\documentclass[a4paper,amsmath,twocolumn,superscriptaddress]{revtex4}

\usepackage{amsmath,amssymb,graphicx}
\usepackage{ mathrsfs }
\usepackage{xcolor}
\usepackage[makeroom]{cancel}
\usepackage[noend]{algpseudocode}
\usepackage{algorithm}

\newcommand{\ket}[1]{\left|#1\right\rangle}
\newcommand{\bra}[1]{\left\langle#1\right|}
\newcommand{\braket}[2]{\left\langle#1|#2\right\rangle}
\newcommand{\avg}[1]{\left\langle#1\right\rangle}
\newcommand{\comm}[2]{\left[#1,#2\right]}

\newcommand{\tH}{\tilde{H}}
\newcommand{\sgn}{\text{sgn}}

\newcommand{\der}[2]{\frac{d\, #1}{d\, #2}}
\newcommand{\dder}[2]{\frac{d^2\, #1}{d\, #2^2}}
\newcommand{\pder}[2]{\frac{\partial\, #1}{\partial\, #2}}

\renewcommand{\Im}{\text{Im}}
\renewcommand{\Re}{\text{Re}}
\newcommand{\cc}{c.c.}
\newcommand{\expT}[1]{\exp_{\mathcal{T}}\left(#1\right)}

\newcommand{\remove}[1]{}

\begin{document}
\title{Behavior of Analog Quantum Algorithms}
\author{Lucas~T.~Brady}
\email{Lucas.Brady@nist.gov}
\affiliation{Joint Center for Quantum Information and Computer Science, NIST/University of Maryland, College Park, Maryland 20742, USA}
\affiliation{Joint Quantum Institute, NIST/University of Maryland, College Park, Maryland 20742, USA}
\author{Lucas~Kocia}
\affiliation{Sandia National Laboratories, Livermore, California 94550, USA}
\author{Przemyslaw Bienias}
\affiliation{Joint Center for Quantum Information and Computer Science, NIST/University of Maryland, College Park, Maryland 20742, USA}
\affiliation{Joint Quantum Institute, NIST/University of Maryland, College Park, Maryland 20742, USA}
\author{Aniruddha Bapat}
\affiliation{Joint Center for Quantum Information and Computer Science, NIST/University of Maryland, College Park, Maryland 20742, USA}
\affiliation{Joint Quantum Institute, NIST/University of Maryland, College Park, Maryland 20742, USA}
\author{Yaroslav Kharkov}
\affiliation{Joint Center for Quantum Information and Computer Science, NIST/University of Maryland, College Park, Maryland 20742, USA}
\affiliation{Joint Quantum Institute, NIST/University of Maryland, College Park, Maryland 20742, USA}
\author{Alexey V. Gorshkov}
\affiliation{Joint Center for Quantum Information and Computer Science, NIST/University of Maryland, College Park, Maryland 20742, USA}
\affiliation{Joint Quantum Institute, NIST/University of Maryland, College Park, Maryland 20742, USA}

\date{\today}
\begin{abstract}
Analog quantum algorithms are formulated in terms of Hamiltonians rather than unitary gates and include quantum adiabatic computing, quantum annealing, and the quantum approximate optimization algorithm (QAOA).  These algorithms are promising candidates for near-term quantum applications, but they often require fine tuning via the annealing schedule or variational parameters.  
In this work, we explore connections between these analog algorithms, as well as limits in which they become approximations of the optimal procedure.
Notably, we explore how the optimal procedure approaches a smooth adiabatic procedure but with a superposed oscillatory pattern that can be explained in terms of the interactions between the ground state and first excited state that effect the coherent error cancellation of diabatic transitions.  Furthermore, we provide numeric and analytic evidence that QAOA emulates this optimal procedure with the length of each QAOA layer equal to the period of the oscillatory pattern.  Additionally, the ratios of the QAOA bangs are determined by the smooth, non-oscillatory part of the optimal procedure.  We provide arguments for these phenomena in terms of the product formula expansion of the optimal procedure. 
With  these  arguments,  we  conclude  that  different  analog  algorithms  can emulate the optimal protocol under different limits and approximations.
Finally, we present a new algorithm for better approximating the optimal protocol using the analytic and numeric insights from the rest of the paper.  In practice,  numerically, we find that this algorithm outperforms standard QAOA and naive quantum annealing procedures.
\end{abstract}

\maketitle

\section{Introduction}

Analog quantum algorithms come in a variety of forms, from Adiabatic Quantum Computing \cite{Farhi2000} and Quantum Annealing \cite{Kadowaki1998} to variational algorithms such as the quantum approximate optimization algorithm (QAOA) \cite{Farhi2014}.  Analog quantum algorithms are particularly relevant in the Noisy Intermediate Scale Quantum device \cite{Preskill2018} era, where they are capable of running effectively on small-scale devices.

All these analog quantum algorithms use the same basic ingredients but combined in different ways that obfuscate the connections between these algorithms.   
Adiabatic Quantum Computing slowly changes the system from an initial Hamiltonian, whose ground state you start in, to a final Hamiltonian, whose ground state you want to know.  By going slowly, Adiabatic Quantum Computing can rely on the adiabatic theorem \cite{Jansen} which ensures this state transfer so long as the ramp is smooth and monotonic and the runtime scales as an inverse polynomial of the spectral gap.
Quantum Annealing is a broader algorithm that allows for non-adiabatic effects, and this has recently led to the field of diabatic quantum annealing \cite{Crosson2020} that explicitly uses excitations above the ground state to solve problems faster, the difficulty being how to control and utilize these excitations.

QAOA is based upon a different, variational framework where the Hamiltonian evolution obeys bang-bang structure at all times. Here, the quantum optimization problem is solved by optimizing the lengths of Hamiltonian pulses in a hybrid, quantum-classical loop. Not much is known about how QAOA relates to other analog quantum algorithms and how its performance scales with the number of variational parameters. While originally motivated by a Trotterization of Adiabatic Quantum Computing and Quantum Annealing, QAOA performs quite differently in practice. The numerical results of \cite{Zhou2018,Pagano2019} show that QAOA variational parameters fall along certain smooth curves as the depth of the circuit increases.  These curves superficially resemble a Trotterization of an annealing path, but the size of the Trotter steps is insensitive to the circuit depth, invalidating standard Trotter error arguments.  There is also numerical evidence \cite{Zhou2018} that these curves, when interpreted as annealing paths, exhibit properties of diabatic speedups.

More recently, techniques from optimal control theory \cite{Pontryagin} have been applied to analog quantum algorithms \cite{Yang,Bapat2018,Mbeng,Lin,Brady2020}, specifically in the context of the variational approach of QAOA.  These optimal control techniques were applied to the more generalized problem of analog quantum algorithms in Ref.~\cite{Brady2020}.  This optimal protocol takes on a bang-anneal-bang form with guaranteed bangs at the beginning and end that become vanishingly smaller as the allowed time for the protocol increases.  In the middle, the protocol often takes on an annealing-like form with a smooth control function.  We refer to this optimized protocol as an optimal curve/protocol throughout the paper.

This paper focuses on analytically proving the connections among all the analog algorithms mentioned previously.  First, we show that QAOA emulates the optimal curve, acting as a large-time-step Trotterization of this curve. Second, we show that, in the limit of long time, the optimal
curve resembles an optimal adiabatic path similar to the annealing schedule of Roland and Cerf \cite{Roland}, which was optimized to ensure adiabaticity with respect to the instantaneous spectral gap.
Therefore, the asymptotic curves discovered for QAOA in Refs.~\cite{Zhou2018,Pagano2019} are derived from the optimal adiabatic path of the system.  In the short time and low circuit depth limit, the optimal 
curve and QAOA are still connected and begin to resemble excited state computation seen in diabatic quantum annealing \cite{Crosson2020}.

These results rely on the fact that the optimal schedules have large annealing-like regions, which \cite{Brady2020} showed are common in optimal curves.  In practical terms, the relationship between QAOA and the optimal curves means that QAOA can safely be scaled up by \emph{bootstrapping}, or using the results of lower circuit depth optimization to produce a good guess for the variational parameters in a higher circuit depth setting.  This bootstrapping method was suggested by Refs.~\cite{Zhou2018,Pagano2019}, and our results in this paper seek to understand why this method is valid.  This relationship also means that QAOA parameters can be used to form an initial ansatz for the optimal schedule, which in general has better performance than QAOA.  Furthermore, our results contradict the common design philosophy that annealing paths should be monotonic (see reverse annealing as a notable exception \cite{Perdomo-Ortiz,Chancellor}).  Monotonicty is a holdover from the infinite-time adiabatic limit \cite{Farhi2000}, and a monotonic schedule improves the energy of the state over doing nothing \cite{Callison}.  However, our results show that adding an oscillation to the annealing schedule, with a frequency dependent on the spectral gap, improve performance by coherently cancelling the error due to leakage to excited states.  The amplitudes of these oscillations vanish in the infinite-time limit.

Finally, we present a new practical algorithm for approximating the bang-anneal-bang optimal control protocol.  This algorithm uses the analytic and numeric results from the rest of the paper to create an ansatz for the form of the optimal protocol.  This ansatz has a small number of variational parameters, and the number of parameters can be taken to be independent of system size or scaled up with system size depending on the available resources.  We demonstrate that this new ansatz outperforms both QAOA and a naive monotonic annealing schedule. 

We begin in Section \ref{sec:algorithms} by reviewing the relevant algorithms and providing background information on them.  Section \ref{sec:numerics} provides the original numerical motivation for this work, presenting the QAOA asymptotic curves of \cite{Zhou2018,Pagano2019} and the numerical connection between these curves and the optimal schedules of \cite{Brady2020}.  In order to prove this connection, our analytics are broken up into two parts.  The first analytic part in Section \ref{sec:oscillations} relates to the optimal curves themselves, showing how the oscillatory behavior arises.  This section explores the properties of the initial and final bangs, which serve to spread the population out into more than just the ground state and then bring it back together, with the intermediate annealing region providing a nearly-adiabatic procedure. 
The oscillations result from properties similar to counter-diabatic driving terms from shortcuts to adiabaticity \cite{Demirplak,Berry,Guery-Odelin}.  The second analytic part in Section \ref{sec:product} presents work involving product-formula expansions.  
We show that, if the underlying annealing curve consists of a smooth slow curve and a fast small oscillation, a product formula of that annealing curve gets a reduction in its error bounds when the product formula step sizes match the period of the oscillations.
This reduction, combined with potential coherent error effects and additional optimization, can help explain the step size of QAOA and how it relates to the optimal curves.  We present our new bang-anneal-bang ansatz algorithm in Section \ref{sec:bab_ansatz}.  Finally, in Section \ref{sec:conc}, we summarize and review the implications and caveats of our work, providing possible directions for future study and development.

\section{The Algorithms}
\label{sec:algorithms}

All of the analog quantum algorithms considered here fit within a linear control framework described by the Hamiltonian
\begin{equation}
    \label{eq:base_Hamiltonian}
    \hat{H}(t) = u(t) \hat{B} + (1-u(t))\hat{C}.
\end{equation}
The Hamiltonian $\hat{B}$ is often described as the ``mixer'' and encodes quantum mixing (e.g.\ a uniform transverse field on qubits).  $\hat{C}$ is known as the ``problem'' Hamiltonian and encodes the optimization task (e.g.\ a diagonal Hamiltonian with the target cost function along the diagonal).  In all examples, the initial state of the system is taken to be the ground state of $\hat{B}$, and the target state of the system is taken to be the ground state of $\hat{C}$.  The control function $u(t): [0,t_f] \rightarrow [0,1]$ specifies the time evolution protocol of the algorithm over its total runtime $t_f$. The analog algorithms studied here each come with a different design ansatz for this control function.

\subsection{Adiabatic Quantum Computing}

Adiabatic Quantum Computing was originally proposed to solve combinatorial optimization problems \cite{Farhi2000}.  The function $u(t)$ is taken to be a monotonic function, starting at $u(0)=1$ and ending at $u(t_f) = 0$.  If the change in $u(t)$ is slow enough, the quantum adiabatic theorem \cite{Jansen} guarantees adiabaticity,  which means that the system will stay in the same relative eigenstate throughout the evolution.  Notably, this is usually employed to ensure that a system starting in the ground state of $\hat{B}$ at $t=0$ will evolve into the ground state of $\hat{C}$ at $t=t_f$.

This necessitates that the Hamiltonian, $\hat{H}(t)$, maintains a non-zero spectral gap throughout, with some exceptions (e.g.~ground state degeneracy for all $t$ or just at $t=t_f$) \cite{Farhi2000}.  A commonly cited condition for adiabaticity is that \cite{Farhi2000}
\begin{equation}
t_f \gg \frac{\left|\left|\der{\hat{H}(t)}{(t/t_f)}\right|\right|}{\min_{t}\Delta(t)^2},
\end{equation}
where $\Delta(t)$ is the spectral gap of the Hamiltonian at time $t$.  This is a simplified condition that often works in practice, and its formal version, while more complicated \cite{Jansen}, depends roughly on the same parameters, potentially with worse exponents.

Therefore, a large part of the analysis of adiabatic quantum algorithms involves spectral theory to determine the size of $\Delta(t)$.  During most of the anneal, the spectral gap is usually independent of $n$, but during avoided level crossings, which often correspond to phase transitions, the gap can close polynomially or exponentially with $n$.  In hard optimization problems, this spectral gap is exponentially small in the vicinity of avoided level crossings.

Often the monotonic annealing schedule, $u(t)$, is taken to be a linear ramp (or some other hardware-determined shape), but the ramp can be optimized to slow down when the gap is small and speed up when the gap is large.  This optimization, originally proposed by Roland and Cerf \cite{Roland}, is necessary to recover the Grover quadratic speed-up for unstructured search, and there is good evidence that optimization of the schedule in general can lead to a similar quadratic speed-up over unoptimized schedules \cite{Jarret2018}.  The optimized Roland and Cerf schedule is specific to the unstructured search problem, but it can be generalized by methods such as the quantum adiabatic brachistochrone \cite{Rezakhani}.  One problem with these optimized schedules is that they require full knowledge of the spectral gap to construct.  The unstructured search problem has the same spectral structure in all problem instances, but knowledge of the spectral gap is generally hard to find \textit{a priori}.  Another problem of such optimized schedules 
is that, if the minimal spectral gap is exponentially small, they might require realistically unachievable exponential precision 
\cite{Hen2018}.

\subsection{Quantum Annealing}

Quantum Annealing was originally proposed \cite{Kadowaki1998} before Quantum Adiabatic Computing and was justified not by the adiabatic theorem, but instead by comparison to classical thermal annealing.  In practice, the setup of Annealing is roughly the same as Adiabatic, with $u(0)=1$, $u(t_f) = 0$, and a smooth, usually monotonic ramp in between.

The relative definitions of Annealing and Adiabatic are slightly ambiguous and vary throughout the field.  In this paper, we will use one of the more common definitions of Quantum Annealing as a generalization of Adiabatic Quantum Computing, with Adiabatic being a subclass of Annealing.  Whereas Adiabatic Quantum Computing requires adiabaticity, meaning the state of the system must always track the ground state, Quantum Annealing allows for either non-adiabatic effects  or adhering to adiabaticity.
These non-adiabatic effects might be due to thermal noise or simply going too fast (in the present paper, we will consider only unitary dynamics, so there will be no thermal noise).
These non-idealities could mean that the final state is an excited state that is deemed good enough for practical purposes.

It is also possible to utilize the sped up behavior and engineer the dynamics to depopulate the ground state and then repopulate it \cite{Muthukrishnan,Brady2017}, utilizing the power of higher excited states for intermediate computational steps.  This is the basis of diabatic quantum annealing \cite{Crosson2020}.  While diabatic algorithms show promise, it is currently unclear how to efficiently engineer the desired effects.  This paper could be interpreted as addressing this question, and we point interested readers to Section \ref{sec:bab_ansatz} where we describe a practical algorithm for engineering a useful diabatic evolution.

\subsection{Quantum Approximate Optimization Algorithm (QAOA)}

While QAOA is sometimes described in the digital quantum circuit framework, it is ultimately an analog quantum algorithm.  The control function is no longer smooth but instead takes on a pulsed, bang-bang form where $u(t)$ can only equal 0 or 1, meaning we are only applying either $\hat{B}$ or $\hat{C}$ but not linear combinations of them.  The original \cite{Farhi2014} formulation of QAOA is best described using unitaries where the system starts in an initial state $\ket{x(0)}$ (the ground state of $\hat{B}$) and ends at a final time, $t_f$, in the state
\begin{equation}
    \ket{x(t_f)} = \left[\prod_{i=1}^{p}e^{-i\beta_i \hat{B}}e^{-i\gamma_i \hat{C}}\right]\ket{x(0)}.
\end{equation}
The (positive) times $\vec{\gamma}$ and $\vec{\beta}$, also known as angles, describe how long to apply each bang, with the label $\gamma$ referring to evolution times under $\hat C$ and $\beta$ referring to evolution times under $\hat B$ by convention. The total runtime for this algorithm is $t_f = \sum_{i=1}^p (\gamma_i+\beta_i)$.  The number of layers in QAOA, $p$, is usually fixed, while the angles $\gamma_i$ and $\beta_i$ are allowed to vary freely.

As a hybrid variational algorithm, QAOA uses a classical optimizer to optimize the angle parameters and a quantum computer to sample and estimate the final energy $\avg{E} = \bra{x(t_f)}\hat{C}\ket{x(t_f)}$.  The goal is to prepare a state that is close to the target state by finding the $\gamma_i$ and $\beta_i$ that minimize $\avg{E}$.

As it was originally proposed, QAOA was conceived as a generalized discretization of Quantum Annealing.  Indeed, a Suzuki-Trotter expansion of an Annealing schedule would result in a bang-bang pattern similar to QAOA.  However, numerical results \cite{Zhou2018, Pagano2019} have shown that the optimal angles do not approach a Trotterization. This is evident from the observation that the ideal bang lengths remain roughly constant as $p$ is increased, whereas under Trotterization, they become vanishingly small as $p\to\infty$.  A key goal of the current paper is to explain this phenomenon and describe this large-$p$ behavior.

\subsection{Optimal Schedules}

In a previous study \cite{Brady2020}, the analog quantum algorithm problem was analyzed through the lens of optimal control theory.  We asked what properties an optimal $u(t)$ must have in order to produce the lowest possible $\avg{E}$ within a given amount of time $t_f$.

The resulting schedule takes on a form with a finite-length $u=0$ bang at the beginning and a finite-length $u=1$ bang at the end.  Our analytics suggested multiple possibilities in the middle, but in all numerics tested (mostly focusing on the Ising model with some additional data for the Heisenberg model), the middle region was dominated by a smooth non-monotonic annealing region.  The form of this annealing schedule was not studied extensively, and the exact shape of this region, as well as a heuristic picture of the evolution, is one of the main contributions of the current paper.

The initial and final bangs in such a bang-anneal-bang procedure are guaranteed to decrease as $t_f$ increases and vanish in the limit $t_f\to\infty$ \cite{Brady2020}.  In fact, this corresponds to recovering the adiabatic limit.  Our results in the current paper can be used to interpret the initial and final bangs as exciting the system into a diabatic annealing regime.

\section{Numerically Comparing Optimal Curves and QAOA}
\label{sec:numerics}

Our main results are inspired by two separate pieces of numerical evidence.  The first is the asymptotic large-$p$ structure of QAOA, as has already been presented in Refs.~\cite{Zhou2018, Pagano2019}.  The second is the asymptotic large-$t_f$ behavior of the optimal curves.  This behavior of the optimal curves was explored partially in the Appendices of the previous paper \cite{Brady2020}, but here we formalize those results and connect them to the behavior of QAOA.

\subsection{QAOA Curves}

One of the primary sources of excitement with QAOA is the ability to predict the $\gamma_i$ and $\beta_i$ from similar problem instances.  
It has been observed \cite{Brandao2018} that QAOA angles give rise to similar performance across similar problem instances.
More relevant for our purposes, when considering a fixed problem instance, the optimized QAOA angles form a certain pattern, and the QAOA protocol approaches an asymptotic continuous limit with increasing number of layers $p$.

Specifically, suppose that the optimal QAOA angles for a given $p$ are given by $\gamma_i$ and $\beta_i$, then we can construct continuous functions $\gamma_p(s)$ and $\beta_p(s)$ for $s\in[0,1]$ such that 
\begin{align}
    \gamma_p\left(\frac{i-1}{p-1}\right) &= \gamma_{i},\\
    \beta_p\left(\frac{i-1}{p-1}\right) &= \beta_{i}.
\end{align}
As $p$ increases, it has been noted numerically \cite{Zhou2018,Pagano2019} that these functions $\gamma_p(s)$ and $\beta_p(s)$ converge to asymptotic functions $\gamma(s)$ and $\beta(s)$ that become independent of $p$ in the limit $p\to\infty$.

These asymptotic curves should not be confused with a simple Suzuki-Trotter expansion of some underlying annealing curve.  In order to guarantee an accurate approximation of a Hamiltonian time evolution by a Suzuki-Trotter product formula, the time steps are required to be of vanishing order. However, the asymptotic curves prescribe angles of constant order, so if they were interpreted as a simple Suzuki-Trotter expansion, the expected error would be non-vanishing as the number of QAOA rounds goes to inifinity. 

It is possible to construct an annealing curve from the asymptotic QAOA curve, as done in Zhou \textit{et al.}~\cite{Zhou2018}. Specifically, they define $u(s) = \frac{\beta(s)}{\beta(s)+\gamma(s)}$ as an annealing curve, which is well-motivated in part because it is commonly seen that $\beta(s)$ is dominant at the beginning and $\gamma(s)$ is dominant at the end (the reason for this is connected to the asymptotic shape of the optimal curves that QAOA is emulating as we discuss in Sec.~\ref{sec:oscillations}).  The resulting annealing curve captured a well-known effect from diabatic quantum annealing, so-called diabatic cascades \cite{Muthukrishnan}, providing an empirical link between QAOA and diabatic quantum annealing.

\begin{figure}
        \begin{center}
                \includegraphics[width=0.5\textwidth]{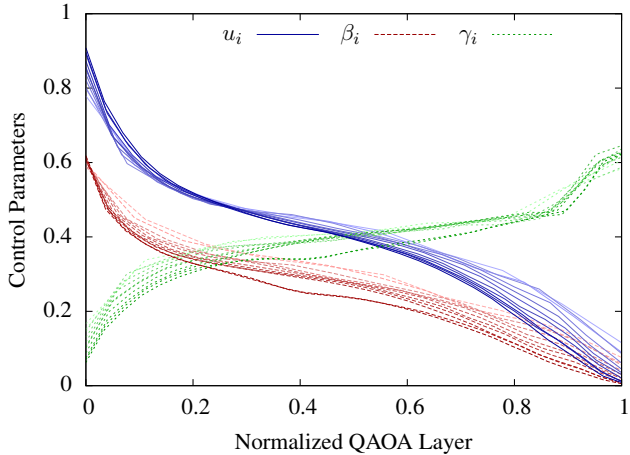}
        \end{center}
        \caption{This plot shows the QAOA variational parameters for a single problem instance at several different values of $p$.  Plotted are the $\gamma_i$, $\beta_i$, and $u_i = \frac{\beta_i}{\beta_i+\gamma_i}$.
        The $x$-axis is the normalized QAOA layer $\frac{i-1}{p-1}$.
        The lighter curves are for lower $p$ (starting at $p=10$), and the darker curves are for higher $p$ (ending at $p=30$).  These curves do vary slightly, but especially at higher $p$, they settle into some smooth asymptotic curve.  This data was gathered for $\hat{B}$ being a transverse field and $\hat{C}$ being a randomized Ising model with all-to-all couplings drawn at random uniformly from the range $[-1,1]$ on $n=8$ qubits (exact couplings given in Appendix \ref{app:Jij}).  }
        \label{fig:qaoa_curves}
\end{figure}

An example of these asymptotic QAOA curves is given in Fig.~\ref{fig:qaoa_curves}.  These numerics indicate that there is some asymptotic curve for each problem instance that QAOA angles are converging to.  Notably, this means that QAOA can bootstrap itself up, using lower $p$ parameters to create good guesses for what the higher $p$ parameters are.

\subsection{Bang-Anneal-Bang Oscillations}

\begin{figure}
        \begin{center}
                \includegraphics[width=0.5\textwidth]{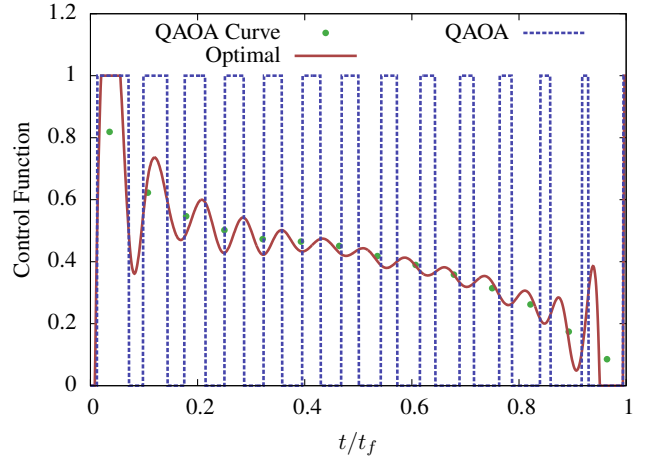}
        \end{center}
        \caption{This plot numerically demonstrates some of the key points of this paper, showing a $p=14$ QAOA protocol and the optimal protocol that takes the same length of time.  These numerics correspond to the same problem instance shown in Fig.~\ref{fig:qaoa_curves}, and the time for the optimal protocol is fixed based on the time taken by the QAOA protocol.  Notice first that the optimal protocol oscillates in such a way that it fits $14$ oscillations into this time frame.  Also, in green, we plot the QAOA variables defined as $u_i = \frac{\beta_i}{\beta_i+\gamma_i}$ which track the underlying annealing portion of the optimal curve (for details see main text).  These properties have been seen numerically in every Ising model we have studied.}
        \label{fig:oscillations}
\end{figure}

The new numerics that inspired this current study involve the bang-anneal-bang behavior of the optimal curves when compared to QAOA. The runtime of a QAOA protocol can be defined in terms of its variational parameters as 
\begin{equation}
    t_f = \sum_{i=1}^p (\gamma_i+\beta_i).
\end{equation}
It is natural then to ask what the optimal curve is for that length of time.  The numeric answer is exemplified in Fig.~\ref{fig:oscillations}.

In Fig.~\ref{fig:oscillations}, we plot the QAOA bangs for a particular instance of a randomized Ising model alongside the optimal curve, with a bang-anneal-bang structure, that takes the same length of time as QAOA.  For ease of optimization, the QAOA instance here uses the same time length, $(\gamma_i+\beta_i)$, for each layer (with that length also being treated as a variational parameter), but all the qualitative properties apply in the normal QAOA setting as well.  Also plotted is the QAOA curve defined by $u_i = \frac{\beta_i}{\beta_i+\gamma_i}$ with these points plotted on the $x$-axis at the midpoint of the corresponding QAOA layer.

There are two key qualitative points to be made here.  First, the optimal curve oscillates about some base curve.  The period of these oscillations matches up with the length of the QAOA layers, with there being $p=14$ QAOA layers and $14$ oscillations of the optimal curve.  Second, the underlying curve that is being oscillated about matches up with the QAOA curve.  These are very general properties and were seen in every numerical instance studied.

This behavior suggests a connection between QAOA and bang-anneal-bang form of the optimal procedures.  Furthermore, when the optimal procedure is given a long time, it approaches an adiabatic procedure, with the initial and final bangs becoming vanishingly small and the amplitude of the oscillations approaching zero.  The rest of the paper will be devoted to explaining this connection by focusing on the two parts of this problem.

In Section \ref{sec:oscillations}, we explain where these oscillations come from, employing an asymptotic near-adiabatic perturbative analysis.  In the long-$t_f$ limit, the period of oscillations turns out to be inversely proportional to the instantaneous spectral gap; although, the smaller $p$ used in current QAOA implementations result in a $t_f$ such that this limit is not reached and the periods do not correspond to the spectral gap.  It could be possible to extract spectral gap information from a long enough QAOA procedure, but that is likely to be outside the regime of near-term quantum computers.  Numerically, the examples we can access also are not in this asymptotic regime, but the same analytic mechanism can explain the origin of these oscillations even if the timescales are not long enough for the spectral gap to govern the oscillation period.

Then in Section \ref{sec:product}, we explain the connection between these oscillations and QAOA by interpreting QAOA as a large-time-step product formula (a.k.a.~Trotterization) of the underlying optimal curve.  Due to the large timescales involved, QAOA cannot ordinarily be interpreted as a product formula without incurring untenable errors.  
We show that a product formula aligning with an underlying oscillation incurs less error overall; though, our method does still have scaling with $p$ that could potentially be mitigated by coherent cancellation of Trotter errors.

Because these properties of the optimal curve are qualitatively universal, we can utilize them to produce an ansatz for the optimal protocol.  We do this in Section \ref{sec:bab_ansatz} and show that this ansatz, which includes only a small number of variational parameters, can outperform both naive annealing and QAOA.

\section{Deriving the Oscillations}
\label{sec:oscillations}

First we consider how to characterize the optimal curve, specifically the oscillatory pattern in the annealing portion of its bang-anneal-bang form.  
The interior annealing region mostly has a smooth annealing form which is quite apparent for transverse field Ising models but appears to varying degrees in other models \cite{Brady2020}.

In the large-runtime limit, these optimal curves become a monotonic annealing schedule, and the oscillations have vanishing amplitude.  This is consistent with the adiabatic theorem \cite{Jansen}, which guarantees that a monotonically decreasing control function will transform an initial ground state to a final ground state.  Interestingly, there are conjectures that, in the space of the Lie algebra generated by $\hat B$ and $\hat C$, the shortest path that transforms between the terminal ground states is precisely the adiabatic path which transfers all eigenstates in the initial Hamiltonian to the equivalent eigenstates in the final Hamiltonian \cite{Bukov}.  Their conjecture was proven in the adiabatic and near-adiabatic limit, but is harder to prove far away from this limit.  Furthermore, the result considers paths that prepare the exact final ground state, which is a valid assumption in our setting only in the limit of long runtimes. Therefore, we expect the optimized annealing schedules for long times to approach an optimized adiabatic schedule, similar to what was derived by Roland and Cerf \cite{Roland} for the unstructured search problem or in the quantum adiabatic brachistochrone \cite{Rezakhani}.

It is possible to emulate an adiabatic protocol in a shorter period of time, using shortcuts to adiabaticity and counter-diabatic protocols \cite{Guery-Odelin}, and most notably, it is even possible to emulate the effects of a CD addition to the Hamiltonian using only the original Hamiltonians with a fast oscillation of the control function \cite{Petiziol}.
These fast oscillations rely on user-defined periods and parameters and so do not describe the properties seen in the numerics for the optimal schedule.
This method also relies on full knowledge of the counter-diabatic driving term which we lack and which is difficult to find for large system sizes.

\begin{figure}
        \begin{center}
                \includegraphics[width=0.5\textwidth]{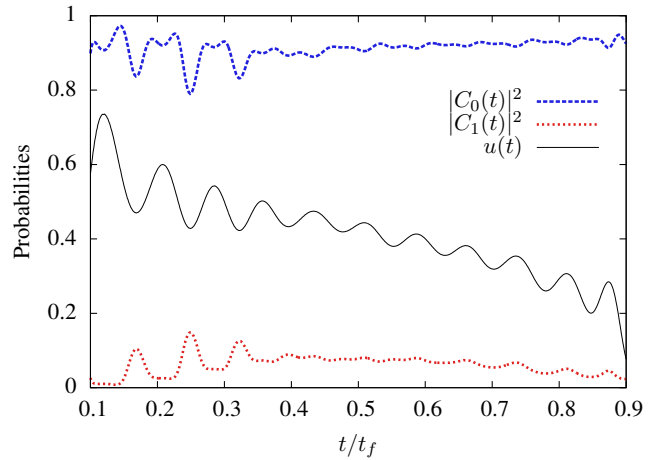}
        \end{center}
        \caption{Here we zoom in on the optimal curve from Fig.~\ref{fig:oscillations} during the annealing region.  In addition, we plot the probabilities of being in the instantaneous ground state, $|C_0(t)|^2$, and first excited state, $|C_1(t)|^2$.  This plot shows that this annealing region is transferring the states adiabatically with the populations roughly maintained from the beginning to end of the anneal.  There is variation in these amplitudes but they roughly return to themselves after a full oscillation of the annealing curve.}
        \label{fig:eigdist}
\end{figure}

Before we proceed, we comment on whether we should expect an adiabatic evolution, or at least one that keeps the instantaneous eigen-distribution constant, potentially only at certain points (at least constant between the beginning and end of the annealing region).  Numerically, we do see this in the optimal curves.  For long times, as stated previously, the anneal is just an optimized adiabatic schedule with very small oscillation amplitudes.  Whereas for shorter times, the oscillations are quite pronounced, and an examination of the eigen-distribution, such as in Fig.~\ref{fig:eigdist} reveals that the instantaneous eigen-distribution does indeed remain relatively constant, matching up at the beginning and end of the anneal as well as at points roughly in line with the periods of the oscillations.  

Therefore, it seems natural that the annealing region of the optimal curves is emulating a shortcut to adiabaticity approach.  The Magnus expansion method of Ref.~\cite{Petiziol} uses the frequency of their oscillations as a fit parameter with their method relying on a perturbative approach as this period becomes small.  Since our goal is to derive the frequency of the oscillations rather than impose a frequency, this Magnus expansion method is not useful in our circumstance.  Here we consider the following different approach to derive such an oscillatory counter-diabatic procedure. 

\subsection{Near Adiabatic Approximation}

To demonstrate this approach, we will first restrict to the setting where only the ground state and first excited state are relevant.  This holds in the near-adiabatic limit and is supported by our numerics, shown in Fig.~\ref{fig:eigdist}.

The methods we use here are similar to those used in adiabatic boundary cancellation methods \cite{Lidar2009,Wiebe2011} and are generally connected to adiabatic analysis \cite{MacKenzie2006,Amin2009} and the analysis of shortcuts to adiabaticity \cite{Guery-Odelin}.  We specifically follow results from \cite{Brady2018}.

Consider a case where we have some control function,
\begin{equation}
    u(t) = u_0(t)+ c(t)\, ,
\end{equation}
so that the Hamiltonian is
\begin{align}
    \hat{H}(t) &= \hat{H}_0(t) + \hat{H}_c(t) 
    \\\nonumber
    &= \left(u_0(t)\hat{B}+(1-u_0(t))\hat{C}\right)+\left(c(t)(\hat{B}-\hat{C})\right)\, .
\end{align}
Here, $u_0(t)$ is a function determined by the adiabatic nature of the problem.  Any sufficiently slow procedure is adiabatic, but the annealing schedule can be optimized, such as the analytic fine-tuning of Roland and Cerf \cite{Roland}, to improve performance and ensure the onset of adiabaticity at smaller $t_f$.  The function $c(t)$ represents our control freedom, and we can choose it so that the adiabatic passage described by $u_0(t)$ is followed as precisely as possible.

We will furthermore label the instantaneous eigenstates of $\hat{H}_0(t)$ by $\ket{j_0(t)}$ with eigenvalues $\lambda_j(t)$ so that
\begin{equation}
    \label{eq:eig}
    \hat{H}_0(t)\ket{j_0(t)} = \lambda_j(t) \ket{j_0(t)},
\end{equation}
and we ignore any degeneracies (to account for degeneracies, we could work in a subspace defined by the symmetries of our Hamiltonian and our initial ground state).  These eigenstates are defined up to a phase choice which will be set below.  Throughout this section (unless otherwise noted), we use the ${}_0$ subscript to indicate that these quantities are relative to the eigenframe determined by $u_0(t)$ rather than the full eigenframe determined by $u(t)$.

Now, we can express our current state in terms of these eigenstates by
\begin{equation}
    \ket{\psi(t)} = \sum_{j} C_j(t)\ket{j_0(t)}.
\end{equation}

We make the assumption that $|C_0(t)|$ and $|C_1(t)|$ are much larger than all other probability amplitudes.  This assumption implies that $\dot{u}_0$ is small so that the system is evolving approximately adiabatically.  Also it implies that $c(t)$ is small so that the small deviations from the base curve also do not break the approximation of a two-level system.

Applying the Schr\"odinger equation produces
\begin{align}
    \label{eq:NA1}
    \sum_j i\left(\der{C_j(t)}{t}\ket{j_0(t)} + C_j(t) \der{}{t}\ket{j_0(t)} \right)
    \\\nonumber
    =\sum_j \hat{H}(t)C_j(t)\ket{j_0(t)}\, .
\end{align}

Using the orthonormality of the eigenstates $\ket{j_0(t)}$ we can reduce this to a system of coupled differential equations.  For instance, the coefficient of the eigenstate $\ket{k_0(t)}$ in the left hand side of the above equation is 
\begin{equation}
        \label{eq:LHS_near_adiabtic}
         i\left(\der{C_k(t)}{t} + \sum_{j\neq k} C_j(t) \bra{k_0(t)}\der{}{t}\ket{j_0(t)} \right)\, ,
\end{equation}
where we have set the phases of the eigenstates by requiring that $\bra{k_0(t)}\der{}{t}\ket{k_0(t)} = 0$. This choice of phase is fairly common in adiabatic analysis and shortcuts to adiabaticity where it is often referred to as part of the adiabatic frame.
To see why this phase can be chosen, consider $0 = \der{}{t}\braket{k_0(t)}{k_0(t)} = (\der{}{t}\bra{k_0(t)})\ket{k_0(t)}+\bra{k_0(t)}\der{}{t}\ket{k_0(t)}$.  Therefore,  $\Re(\bra{k_0(t)}\der{}{t}\ket{k_0(t)}) = 0$ is always automatically satisfied, while the phase of the state can always be chosen such that $\Im(\bra{k_0(t)}\der{}{t}\ket{k_0(t)}) = 0$.

If the Hamiltonian were stoquastic (Hamiltonians where the off-diagonal elements are all real and non-positive), this choice of phase would mean that the instantaneous ground state maintain the same phase throughout the evolution, which we take to be real and positive.  Similarly, the stoquastic first excited state can be represented using only real amplitudes throughout.

The instantaneous eigenvalues of the Hamiltonian are defined by Eq.~\ref{eq:eig}, and we set $\lambda_0(t)=0$ at all times without loss of generality.
The time derivative of Eq.~\ref{eq:eig} yields
\begin{align}
    &\der{\hat{H}_0(t)}{t}\ket{j_0(t)}+\hat{H}_0(t)\der{}{t}\ket{j_0(t)}\\\nonumber
    &=\der{\lambda_j(t)}{t}\ket{j_0(t)}+\lambda_j(t)\der{}{t}\ket{j_0(t)}.
\end{align}
The inner product of this equation with another eigenstate $\ket{k_0(t)}$ such that $k\neq j$:
\begin{align}
    &\bra{k_0(t)}\der{\hat{H}_0(t)}{t}\ket{j_0(t)})+\bra{k_0(t)}\hat{H}_0(t)\der{}{t}\ket{j_0(t)}
    \\\nonumber
    &=\der{\lambda_j(t)}{t}\braket{k_0(t)}{j_0(t)}+\lambda_j(t)\bra{k_0(t)}\der{}{t}\ket{j_0(t)}.
\end{align}
We can act on the bra states with the Hamiltonian and eliminate one element through orthogonality to get
\begin{align}
    \frac{\bra{k_0(t)}\der{\hat{H}_0(t)}{t}\ket{j_0(t)})}{(\lambda_j(t)-\lambda_k(t))}
    =\bra{k_0(t)}\der{}{t}\ket{j_0(t)}.
\end{align}

In Eq.~(\ref{eq:LHS_near_adiabtic}),  this time derivative of eigenstates is multiplied by the amplitudes $C_j(t)$.  By our assumptions, only $C_0(t)$ and $C_1(t)$ will be relevant, and we can discard cases where $j\neq0,1$.

Now consider the right-hand side of Eq.~(\ref{eq:NA1}).  If we were in the true adiabatic reference frame of the full $\hat{H}(t)$ instead of just $\hat{H}_0(t)$, the Hamiltonian would just scale each eigenstate by its eigenvalue, but instead we get
\begin{align}
    &\bra{k_0(t)}\hat{H}(t)\sum_{j}C_j(t)\ket{j_0(t)}
    \\\nonumber
    &= C_k(t)\lambda_k(t)+\sum_j C_j(t)\bra{k_0(t)}\hat{H}_c(t)\ket{j_0(t)}.
\end{align}

We define
\begin{align*}
    \gamma(t)
    &\equiv
    \bra{0_0(t)}(\hat{B}-\hat{C})\ket{1_0(t)}),
    \\
    \Delta(t) &\equiv \lambda_1(t),
    \\
    \kappa_i(t) &\equiv \bra{i_0(t)}(\hat{B}-\hat{C})\ket{i_0(t)},
\end{align*}
where $\lambda_0(t)=0$.  Thus, $\Delta(t)$ has a meaning of the instantaneous spectral gap for the Hamiltonian $H_0(t)$. 
In the stoquastic setting, all of these quantities are real, and we will treat them as such going forward.  

Putting everything together, the Schr\"odinger equation for the ground state and first excited state amplitudes give
\begin{align}
    &i\left(\der{C_0(t)}{t} + C_1(t) \frac{\gamma(t)\dot{u}_0(t)}{\Delta(t)} \right)\label{hA}\\\nonumber
    &= c(t)\left(C_0(t)\kappa_0(t)+C_1(t)\gamma(t)\right),
\end{align}
\begin{align}
    &i\left(\der{C_1(t)}{t} - C_0(t)\frac{\gamma(t)\dot{u}_0(t)}{\Delta(t)}\right)\label{hB}\\\nonumber
    &=\Delta(t)C_1(t)+c(t)\left(C_0(t)\gamma(t)+C_1(t)\kappa_1(t)\right).
\end{align}

With these equations, we can separate out the amplitudes and phases so that
\begin{align}
 C_i(t) = A_i(t)e^{i\varphi_i(t)}.
\end{align}

Separating the real and imaginary parts of the differential equations, the resulting differential equations are
(suppressing all functional dependencies for brevity)
\begin{align}
    \label{eq:na_na_phase}
    \dot{\varphi} 
    &\equiv \dot{\varphi}_0-\dot{\varphi}_1=
    \Delta+c(\kappa_0-\kappa_1)\\\nonumber
    &+\frac{A_0^2-A_1^2}{A_0 A_1} \left(c\gamma \cos(\varphi)-\frac{\gamma \dot{u}_0}{\Delta}\sin(\varphi)\right),
    \\
    \label{eq:na_na_A0}
    \dot{A}_0 &= -\left(c\gamma\sin(\varphi)+\frac{\gamma \dot{u}_0}{\Delta}\cos(\varphi)\right)A_1,
    \\
    \label{eq:na_na_A1}
    \dot{A}_1 &= \left(c\gamma\sin(\varphi)+\frac{\gamma \dot{u}_0}{\Delta}\cos(\varphi)\right)A_0.
\end{align}

As we already mentioned, the assumption that the Hamiltonian is stoquastic 
ensures that the $\gamma$ and $\kappa_i$ functions are real.  In the absence of stoquasticity, these functions could be complex-valued which would have just made the algebra above to separate our amplitudes and phases slightly more complicated without fundamentally changing the results.

The equations (\ref{eq:na_na_A0}, \ref{eq:na_na_A1}) for the amplitudes $A_{0,1}$ can be integrated to give
\begin{align}
    \label{eq:A0_non_adi}
    A_0(t) &= a\cos\left(
        \int_0^t dt'\left(c\gamma\sin(\varphi)+\frac{\gamma\dot{u}_0}{\Delta}\cos(\varphi)\right)+\vartheta
    \right),\\
    \label{eq:A1_non_adi}
    A_1(t) &= a\sin\left(
        \int_0^t dt'\left(c\gamma\sin(\varphi)+\frac{\gamma\dot{u}_0}{\Delta}\cos(\varphi)\right)+\vartheta
    \right),
\end{align}
where $a$ and $\vartheta$ are constants that can be set such that $a\cos\vartheta$ is the initial population of the ground state and $a\sin\vartheta$ is the initial population of the first excited state.  The signs here do not matter because any sign can be absorbed into the $\varphi$ phase.

Maintaining the same populations of $|A_0(t)|$ and $|A_1(t)|$ throughout the evolution translates to the trig argument in Eqs.~(\ref{eq:A0_non_adi})\ \& (\ref{eq:A1_non_adi}) at time $t_f$,
\begin{equation}
    \label{eq:Theta}
    \Theta_0[u(t)] = \int_0^{t_f} dt\left(c\gamma\sin(\varphi)+\frac{\gamma\dot{u}_0}{\Delta}\cos(\varphi)\right),
\end{equation}
being close to a multiple of $\pi$.  However in practice, a non-zero multiple of $\pi$ would correspond to swapping the populations back and forth during the anneal which is inconsistent with the assumptions we made about being near-adiabatic with low leakage.  Therefore, we want $\Theta_0$ to be as close to zero as possible, meaning that the problem has simplified to finding the $c(t)$ that ensures $\Theta_0\approx0$.

In the numerical examples shown, the oscillations fit neatly into the time allowed, giving an integer number of oscillations.  This is largely because we look at cases where the time for the optimal procedure is the same as the time that QAOA takes.  In other cases when the time does not match the time from a QAOA protocol, the oscillations are not regular.  The point here is that we only expect $c(t)$ to have a nice, simple oscillatory structure when $t_f$ neatly divides into periods of the oscillations.

In Appendix \ref{app:NA_pert}, we work in the perturbative limit of $\dot{u}_0\ll1$ to derive $c(t)$ which ensures that $\Theta$ is zero:
\begin{equation}
\label{eq:c_perturbative_limit}
    c(t) = \frac{\dot{u}_0(t)^2}{\Delta(u_0(t))^2}\der{\ln\left(\frac{\Delta(u_0(t))^2}{\gamma(u_0(t))}\right)}{u_0(t)}\cos(\Delta(u_0(t)) t) + \mathcal{O}(\dot{u}_0^3).
\end{equation}
The cosine here follows the oscillations of the phase, $\varphi(t)$, and mean that we go faster when the phase difference is large and slower when the phase difference is small.  Essentially, these oscillations are designed to take advantage of the natural phase oscillations to speed up the procedure.  The dependence of the amplitude on the gap reflects the fact that adiabaticity is easier (and therefore these oscillations are not necessary) when the gap is large.
So long as $t_f$ is a multiple of the period of oscillations $\tau = \frac{2\pi}{\Delta(u_0(t))}+\mathcal{O}(\dot{u}_0(t))$, then these oscillations will ensure that the amplitudes follow the eigenbasis associated with $u_0(t)$ up to corrections of $\mathcal{O}(\dot{u}_0^3)$.

As a note here, it is well known that oscillations, such as these, can eliminate the asymptotic nature of adiabatic theorem \cite{Amin2009,Schiff}.  If oscillations are present, then in the infinite-time limit, the system will no longer be in the ground state, no matter how slowly it evolved.  However, this is not a problem for us because the amplitude of the oscillations is decreasing with $\dot{u}_0^2$ (and hence with $1/t_f^2$) which is small enough for the deleterious effects to not manifest \cite{Amin2009}.

As shown in the Appendix, this perturbative expression relies on the cancellation that occurs when $c(t)$ is out of phase with the oscillations of the phase difference $\varphi(t)$.  Specifically, $c(t)$ and $\cos(\varphi(t))$ need to be in phase (up to integer multiples of $\pi$) to ensure cancellation.  
This can be seen by looking at Eq.~(\ref{eq:Theta}) where having $c(t)\propto\cos(\varphi(t))$ maximizes the effect of the $c(t)$ term, in the perturbative limit, allowing us to cancel out more of the contributions from the $\dot{u}_0$ term.  
Later in Eq.~(\ref{eq:Theta2}), we can see that having oscillations in the control field that change like $\cos(\varphi(t))$ (differentiating to a $\sin(\varphi(t))$) will counteract the $\cos(\varphi(t))$ in that integrand leading to a smaller contribution to the total integral.  Up to first order in $\dot{u}_0$, the phase difference scales like $\varphi(t) = \Delta(u_0(t))t$, so we wind up with oscillations with period inversely proportional to the spectral gap.

Unfortunately, the numerics shown in Figs.~(\ref{fig:qaoa_curves})-(\ref{fig:eigdist}) for the optimal curves do not fall into the perturbative regime described above.  In these numerics, $\dot{u}_0$ is large enough ($t_f$ is small enough) that $\dot{\varphi}$ is no longer dominated by the spectral gap and begins oscillating at a higher frequency.  As stated in the numerics section, determining the exact optimal curves for larger $t_f$ becomes unfeasible due to the difficulty of determining the gradient when so many solutions are good up to numerical precision.

While it is not possible to solve for this frequency perturbatively any more, based on the  analysis in Appendix \ref{app:NA_pert}, as well as numerical simulations of Eqs.~(\ref{eq:na_na_phase}-\ref{eq:na_na_A1}) outside of the perturbative regime of $\dot{u}_0$, the best way to follow adiabaticity along $u_0(t)$ still relies on the phase $\varphi(t)$.  In this nonperturbative regime, it becomes easier to deal with the full adiabatic reference frame that follows the eigenstates of the full $u(t)$.

In Appendix \ref{app:NA_AF}, we derive the near-adiabatic differential equations again, this time following the full control function $u(t)$ instead of $u_0(t)$.  In this setting, we need to impose boundary conditions that $u(0) = u_a$ at the start of the relevant region and that $u(t_f) = u_b$ at the end of the region.  The resulting equations are similar, and the key quantity is still given by an argument that is functionally similar to $\Theta_0$:
\begin{equation}
    \label{eq:Theta2}
    \Theta[u(t)] \equiv \int_0^{t_f} dt\frac{\gamma(u(t))\dot{u}}{\Delta(u(t))}\cos(\varphi(t)).
\end{equation}
Unfortunately, this form does not lend itself to a perturbative approach anymore because the small quantity $\dot{u}$ now contains information about both the base curve and the oscillations.  Fortunately, this form makes it even more clear how to ensure that this quantity should be close to zero.  Namely, by roughly having $\frac{\gamma(u(t))\dot{u}}{\Delta(u(t))}\propto \sin(\varphi(t))$ 
we ensure that the integrand in Eq.~(\ref{eq:Theta2}) changes sign with a frequency twice that of $\varphi(t)$.  This allows the integral to cancel itself out over the oscillations, resulting in a $\Theta[u(t)]$ which is small.
This intuition coincides perfectly with the numerics in the problem.

\begin{figure}
        \begin{center}
                \includegraphics[width=0.5\textwidth]{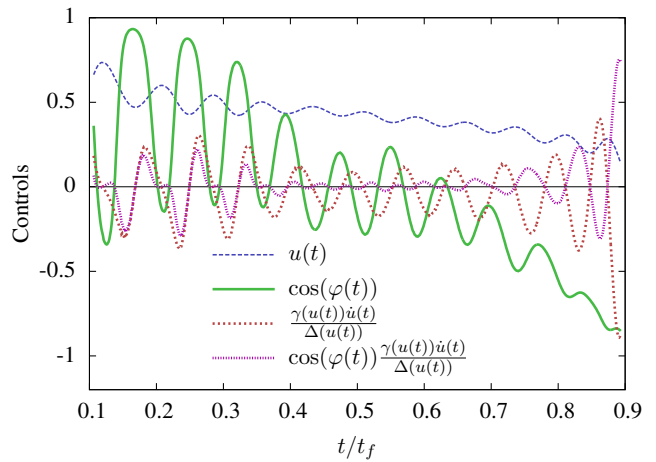}
        \end{center}
        \caption{This plot shows the optimal curve, $u(t)$, as well as two of the quantities that go into Eq.~(\ref{eq:Theta2}).  This plot uses the same problem instance from Fig.~\ref{fig:qaoa_curves}, and the initial and final bits of time have been cut off to focus on just the annealing region of interest.  Notably, from this curve, we can see that the phase difference between the ground and first excited state matches up exactly with the oscillatory pattern of $u(t)$ (with a $\pi$ phase shift) and is out of phase ($\pi/2$ phase shift) with $\dot{u}(t)$ as we expect from the analytic arguments surrounding Eq.~(\ref{eq:Theta2}).
        }
        \label{fig:Theta}
\end{figure}

To see this in practice, in Fig.~\ref{fig:Theta}, we plot $u(t)$, $\cos(\varphi(t))$, and $\frac{\gamma(u(t))\dot{u}}{\Delta(u(t))}$ as determined numerically for an optimal curve resulting from a model using a randomized Ising model.  This plot was constructed so that the time allotted for the annealing curve corresponds to the amount of time that $p=10$ QAOA needed.

In Fig.~\ref{fig:Theta}, the frequency of the phase oscillations matches the frequency of the optimal curve.  
Note that this is still for relatively low $t_f$ where the annealing curve has relatively large amplitude oscillations, meaning the resulting oscillations are not exactly sinusoidal in shape and the period does not exactly mesh up with the asymptotic expectation of the spectral gap.

\section{Product Formula Error}
\label{sec:product}

Based on the numerics presented in Section \ref{sec:numerics}, we see that QAOA is emulating the behavior of the optimal curve that takes the same amount of time.  This section will seek to elucidate how QAOA can emulate the optimal curve, discussed in the previous section, despite the step sizes being large enough to throw off the usual error analysis of product formulas, also known as Trotterization.

\begin{figure}
        \begin{center}
                \includegraphics[width=0.5\textwidth]{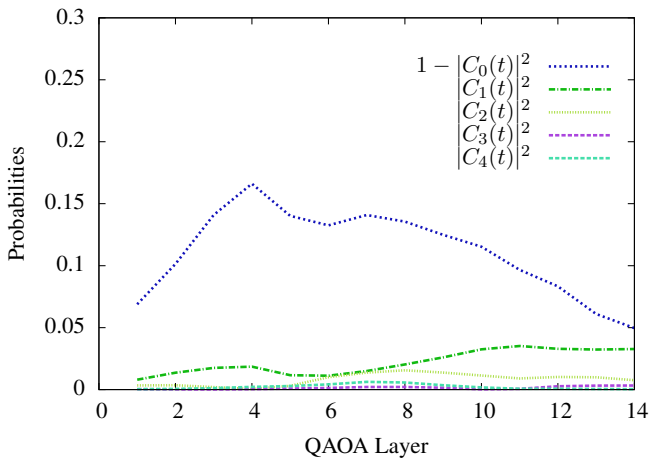}
        \end{center}
        \caption{The instantaneous ground state and first excited state probabilities are plotted versus the QAOA layer.  The instantaneous eigenbasis is defined based off $u_i = \frac{\beta_i}{\beta_i+\gamma_i}$ at the end of the $i$th layer.  The probabilities are measured after the full QAOA layer (both the $\hat{C}$ and the $\hat{B}$ bangs).  QAOA roughly follows an adiabatic-like procedure with the ground state population mostly being preserved.  The problem instance displayed here is the same as in Fig.~\ref{fig:qaoa_curves}.}
        \label{fig:eigdist_qaoa}
\end{figure}

To further see how QAOA is emulating the optimal curve, compare Figs.~\ref{fig:eigdist} \& \ref{fig:eigdist_qaoa}.  These show the probabilities of being in the ground state and first excited state of the instantaneous evolutions for the optimal curve and QAOA, respectively.  For QAOA, the ``instantaneous eigenbasis'' is determined by $u_i = \frac{\beta_i}{\beta_i+\gamma_i}$, the proportion of the QAOA layer dedicated to $\hat{B}$.  This eigenbasis is not physically related to QAOA, which still consists of large bangs, but it catches the effective Hamiltonian being emulated by the pairs of bangs.  Both procedures roughly track the ground state with some leakage, mostly into the first excited state.  QAOA is a rougher procedure with more leakage, compared to the optimal protocol.  Figure \ref{fig:eigdist_qaoa} shows QAOA probabilities only after a full layer of bangs; the intermediate probabilities deviate even more from an adiabatic transfer.

Let's suppose that there is some optimal curve, $u(t)$, defined such that the evolution governed by Eq.~(\ref{eq:base_Hamiltonian})
brings the state as close to the target state as possible in time $t_f$.  Based off the previous section, we will here assume that this optimal control function can be described approximately by
\begin{equation}
    \label{eq:utwsine}
    u(t) = u_0(t/t_f) + c(t,t_f),
\end{equation}
where $c(t)$ is some oscillatory function.  For concreteness, we take
\begin{equation}
    c(t,t_f) = - c_0(t_f) \sin\left(\frac{2\pi}{\tau} t + \phi\right),
\end{equation}
where $c_0$ is some amplitude, $\tau$ some period, and $\phi$ some phase.  We have included the negative sign and specified down to sine since this will correspond to $\phi=0$ later on. 
In essence, we have oscillations whose pattern matches our preexisting pattern of switchings in QAOA protocols and also the pattern of oscillations in the optimal protocols.

For the purposes of this section, we focus on a small region of the annealing curve where $\dot{u}_0$ is small and approximately constant.  Then we ask how accurate a product series approximation is to the true evolution.

The actual evolution will be governed by the unitary time evolution operator,
\begin{equation}
    \label{eq:unitary_timeevo}
    \hat{U}(t_f,0) = \expT{-i\int_0^{t_f}dt\,\hat{H}(t)},
\end{equation}
where $\exp_{\mathcal{T}}$ denotes the time ordered exponential.  We approximate this unitary by breaking it up into a product formula that has a QAOA-like format,
\begin{equation}
    \label{eq:unitary_product}
    \hat{U}_{PF}(t_f,0) = \prod_{k=0}^{p-1}\hat{U}_{1}\left(k\Delta t+\Delta t,k\Delta t\right),
\end{equation}
where $\Delta t$ is the length of the QAOA layer.  The $\hat{U}_1$ here are operators corresponding to a single Trotter slice of the evolution.  In this section, we assume that every QAOA layer uses the same $\Delta t$, and as we will see, this corresponds to the frequency of oscillations in the optimal curve being constant with time (again assuming that $\dot{u}_0$ is small enough).  Therefore, we simply set $\Delta t = \frac{t_f}{p}$.  
It is appropriate to interpret this section as looking at a small region of the optimal curve in the adiabatic limit where the oscillations occur on much shorter timescales than the gross changes in the curve.  When we specify down to our specific control problem, we get
\begin{align}
\label{eq:trotterizedevolution}
    \hat{U}_1\left(t_0+\Delta t, t_0\right) =& \exp\left(-i\hat{B}\int_{t_0}^{t_0+\Delta t} dt\,u(t)\right)
    \\\nonumber
    &\times
    \exp\left(-i\hat{C}\int_{t_0}^{t_0+\Delta t} dt\,(1-u(t))\right).
\end{align}

Our core result is that taking $\Delta t = \tau$, the size of the Trotter slice equal to the period of the annealing oscillation, while keeping the ratio of the bang lengths proportional to $u_0(t)$, leads to a smaller upper bound on the Trotterization error than if we picked a different $\Delta t$.  In this way it becomes advantageous for QAOA to match its layer length to the period of the optimal curve oscillations and its ratio of $\hat{B}$ bang lengths and $\hat{C}$ bang lengths to the value of the base annealing function $u_0(t)$.

We examine this enhancement in two different settings described in the appendices.  First, in Appendix \ref{app:trot_error_op}, we show this enhancement in the context of the standard operator error for product formulas.
The main arguments in the appendix center around the standard error formula for the Trotter approximation known as the product formula.  We zoom in on a small region of the annealing curve and consider the control function given in Eq.~(\ref{eq:utwsine}) vs.\ the case without the added oscillations.  We find that these added oscillations can be accounted for in the product formula, and if the oscillation period and phase match up, it can lead to a lower error bound.  If the oscillation period does not match, the error bound numerically matches up with the error from the case without oscillations at all.

\begin{figure}
        \begin{center}
                \includegraphics[width=0.5\textwidth]{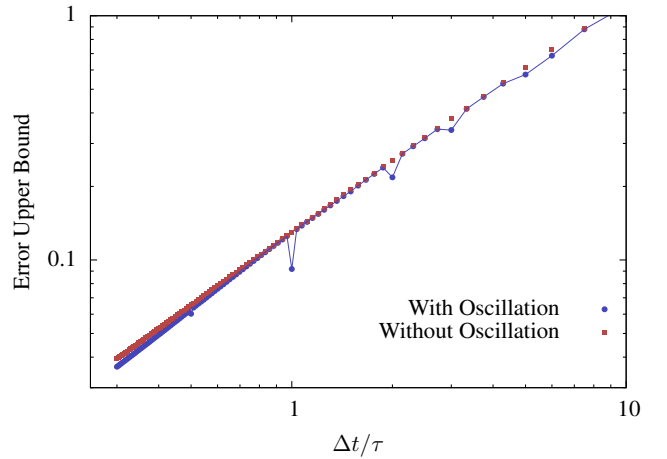}
        \end{center}
        \caption{This plot shows the upper bound on the Trotter error in the unitaries from Eq.~(\ref{eq:Trotter_Product_final}) for a fixed $t_f$, changing $p$ (and hence $\Delta t = \frac{t_f}{p}$).  The oscillation period in this case was taken to be $\tau=0.2$, and the annealing function was taken to be a simple linear ramp with superposed oscillatory function.  The blue dots with lines represent the case with oscillations, and the red squares represent the error bound when no oscillations are present.
        Eq.~(\ref{eq:Trotter_Product_final}) is specifically for the case when $\Delta t = \tau$ with this plot being more general.  As can be seen, the proportional amount of enhancement is greatest when $\Delta t = \tau$, but there is a weaker enhancement when there is an integer multiple difference between these two quantities.}
        \label{fig:trotter_error}
\end{figure}

We find (see the Appendix \ref{app:trot_error_op} for details)  that
\begin{align}
    \label{eq:Trotter_Product_final}
    &||\hat{U}(t_f,0)-\hat{U}_{PF}(t_f,0)||
    \\\nonumber
    &\leq
    \left|\left|\comm{\hat{B}}{\hat{C}}\right|\right|
    \frac{\Delta t^2 p}{2}
    \left(1-\frac{c_0}{\pi}\right),
\end{align}
where we assumed $\Delta t = \tau$ and $\phi = 0$.
In the case of $c_0=0$, this is equal to the standard error bound for product formulas.  This improvement decreases when $\Delta t \neq \tau$, so the enhancement is specifically dependent on matching the size of the Trotter steps to the period of the annealing oscillations.  This washing out can be seen in Fig.~\ref{fig:trotter_error}, but notice that there is also a small enhancement if $\Delta t = m\tau$ or $m \Delta t = \tau$ for any $m\in\mathbb{Z}^+$, $m>1$.

Unfortunately, this bound on the error from unitaries scales linearly with $p$, the number of QAOA layers.  Therefore, rather than getting tighter with more QAOA layers, as we expect, the bound gets looser.  This scaling is because this worst-case bound assumes that adjacent layers of the product formula have errors that accumulate coherently.

Our second approach is restricted to an adiabatic anneal where the goal is to maintain the populations of eigenstates, specifically the ground state in our setting.  The overall Trotter error bound in this setting was recently tightened by Yi and Crosson \cite{Yi2021}.  The same oscillatory enhancement found in the case of operator errors can be shown to occur in this setting as well, but the method requires a perturbative limit which does not hold for the QAOA angles.  Specifically, the method requires $\Delta t \in\mathcal{O}(n^{-1})$ which is not consistent with what we see numerically from QAOA with step sizes remaining roughly constant  and large as the system size $n$ increases, $\beta_i, \gamma_i \in \mathcal{O}(1)$.  We rederive and extend the previous results and modify them for our setting in Appendix \ref{app:trot_error_adi}.  This extension consists of considering the setup where the underlying annealing schedule takes on the form of Eq.~(\ref{eq:utwsine}).  This oscillatory annealing schedule is accounted for in the context of the adiabatic product formula analysis.  The key result of the method of Ref.~\cite{Yi2021} is a reduction in the scaling of the Trotter error from $\mathcal{O}(\Delta t^2 p)$ down to $\mathcal{O}(\frac{1}{p})+\mathcal{O}(\frac{\Delta t}{p})$ when trying to simulate an adiabatic evolution.  We show that this error scaling does not vanish when an oscillating schedule is considered (for small enough oscillations that the adiabatic theorem still holds) and show that there is an enhancement to the error scaling when the product formula step size matches the oscillation period.

It is not possible to fully apply this second approach to our setting because of the perturbative $\Delta t$ issues.  
Our product formula enhancement works partially in this setting and inherits the improved $p$ scaling that the adiabatic Trotter method~\cite{Yi2021} naturally has over  the operator error scaling, Eq.~(\ref{eq:Trotter_Product_final}).

These two approaches are limited to unoptimized product formula approximation of the underlying optimal curve.  Of course, QAOA has more freedom than this and can modify the parameters to do a smarter approximation than just a product formula.  It is allowed to modify the angles away from what a product formula would do in order to achieve more enhancement.  Specifically, it could be possible to coherently match up the leakage between multiple QAOA layers.  All the upper bounds described above assume a worst case scenario that assumes the errors from adjacent QAOA layers add coherently via the triangle inequality, but it may be possible to design the protocol so that the errors subtract coherently to some degree.  Such an approach has been proposed recently \cite{Wurtz} where the Trotterization error itself is used to engineer counter-diabatic driving terms.

We note that the results in this section should all be taken as analytic evidence supporting the numeric evidence from Sec. \ref{sec:numerics}.  
These bounds do exhibit an enhancement when we match up the Trotter step size and the oscillation period, but the bounds are not tight enough to describe the exact setting we see in the numerics.  We leave it up to future work to tighten these bounds further to the setting of QAOA.

\section{Bang-Anneal-Bang Ansatz Algorithm}
\label{sec:bab_ansatz}

One of the leading problems with the optimal curves is how to construct them efficiently.  These optimal curves always seem to have the same qualitative structure, but working out the exact shape and length of various features is key to implementing these schedules effectively.  Formally, these schedules can be found via a gradient descent procedure, using the analytically constructed gradient $\Phi(t) = \frac{\delta \avg{E(t_f)}}{\delta u(t)}$.  This requires information from experimentally inaccessible intermediate times, and numerically estimating this gradient can prove cumbersome for an entire continuous function.

To address these issues, we here present a variational algorithm that produces a good approximation of the bang-anneal-bang optimal path.  This algorithm will not produce the exact optimal procedure but will approximate it, and in our numerical trials it produces better results, given fixed time, than either QAOA or a simple, linear annealing schedule.  The number of variational parameters can be adjusted depending on the available resources.

This algorithm is based off the insight that the asymptotic curve derived from QAOA angles coincides with the base curve in the annealing region of the optimal curves.  Specifically, if QAOA is parameterized in terms of $p$ layers with $\beta_i$, the angles for mixer $\hat{B}$ bangs, and $\gamma_i$, the angles for problem $\hat{C}$ bangs, then the asymptotic QAOA curve can be found in the large $p$ limit by
\begin{equation}
        v\!\left(\frac{i-1}{p-1}\right) = \frac{\beta_i}{\beta_i+\gamma_i},
\end{equation}
where $v(s)$ has the meaning of the control function.
This behavior was noted numerically in Refs.~\cite{Zhou2018,Pagano2019}.  The current work provides justification for the existence of these asymptotic curves and links them to the optimal protocols.  Specifically, \cite{Zhou2018} interpreted this $v(s)$ as an annealing curve which resulted in a good annealing procedure that actually captured well-known effects from diabatic quantum annealing.  Our current algorithm is an improvement on this that captures even more of the structure and power of the optimal procedure.

In optimal protocols, this $v(s)$ has roughly the same functional form as the base curve $u_0(s)$ that determines the shape of the annealing region, up to a superposed oscillatory pattern.  Furthermore, the period of that oscillatory pattern coincides with the duration of the QAOA layers.

Therefore, it should be possible to use an existing QAOA procedure to get a good guess as to what the optimal procedure should look like.  The initial and final bangs are vanishingly small for longer procedures and so are not well captured by QAOA.  These bangs can be inserted in later as variational parameters.  Therefore, we propose the following hybrid variational algorithm for approximating the optimal curves.

\begin{enumerate}

        \item Find QAOA angles for large enough $p$ to be able to identify the shape of $v\left(\frac{i-1}{p-1}\right)$.  In practice, we have found that at $p\sim 5$ it is already possible to  start identifying the pattern, with $p\sim 10-20$ clearly identifying the pattern.
        
        \item Interpret the QAOA curve   $v\left(\frac{i-1}{p-1}\right)$ as a smooth annealing region.
        
        \item Create an ansatz for the bang-anneal-bang curve that has a $u=0$ bang at the beginning, the annealing curve defined by $v\left(\frac{i-1}{p-1}\right)$ in the middle, and a $u=1$ bang at the end.  Furthermore, superpose an oscillatory curve $c(t) = A(t) \sin(\omega(t) t+\phi)$ in the annealing region so that this region is described by $v(t)+c(t)$.
        \begin{itemize}
                \item The lengths of the initial, $\tilde{\gamma}$, and final, $\tilde{\beta}$, bangs are variational parameters.
                \item There are multiple ways to parameterize the interior anneal:
                \begin{itemize}
                        \item The length of the anneal can be fixed to be the same as the time the QAOA procedure took minus the bang lengths, $T_{QAOA}-\tilde{\gamma}-\tilde{\beta}$; the frequency of oscillations can be chosen to be $\omega(t) = 2\pi p/T_{QAOA}$; and the amplitude of the oscillations $A(t) = A$ is taken to be a variational parameter.
                        \item  The length of the anneal can be fixed to be $T_{QAOA}-\tilde{\gamma}-\tilde{\beta}$; the  frequency of oscillations $\omega(t) = \omega$ and amplitude of oscillations $A(t)=A$ are taken to be static variational parameters.
                        \item  The length of the anneal can be fixed to be $T_{QAOA}-\tilde{\gamma}-\tilde{\beta}$; the frequency of oscillations $\omega(t)$ is chosen to be a variable function so that the period of a given oscillation matches the length of the corresponding QAOA layer. The amplitudes of oscillation can either be fixed to be the same or treated as seperate variational parameters in each oscillation.
                        \item The length of the anneal, $T_{A}$, can be treated as a variational parameter~\footnote{If the length of the anneal, $T_A$ is treated as a variational parameter, the algorithm will likely find a local minimum.  Our analytics indicate that the underlying annealing curve will be the same shape regardless of the number of QAOA layers, $p$, or the running time $t_f$.  Therefore, increasing the guess for $T_A$, keeping everything else the same, should still be valid and will likely put your optimization into another local minimum, corresponding to a longer and more accurate procedure.  Assuming the original QAOA had enough parameters to accurately estimate the annealing curve, this can be a good way to increase the accuracy of the procedure without introducing additional variational parameters. }.
                        The frequency can be taken as fixed $\omega(t) = 2\pi p/T_{QAOA}$ or allowed to vary as a free fitted parameter as in previous versions.  The amplitude of oscillation is a single variational parameter or can be binned into different regions with the amplitude in each region being treated as a variational parameter.
                        \item Adjust this ansatz as suits the system at hand and how many variational parameters the specific setting is capable of handling.
                \end{itemize}
                \item Based on analytics, the optimal phase $\phi$ should be $0$, but for optimization purposes it might be beneficial to treat this phase as a variational parameter as well.
                \item In the end, this procedure will result in an ansatz with at least three ($\tilde{\beta}$, $\tilde{\gamma}$, and $A$), but possibly more, variational parameters.
        \end{itemize}
        
        \item  Using the constructed anstaz, run a variational algorithm to determine the optimal values of the selected variational parameters, attempting to optimize with respect to the final energy of the state with respect to $\hat{C}$.

\end{enumerate}

This procedure will always produce a better protocol than just interpreting $u(s)=v(s)$, and the number of variational parameters can be small.  The most intensive part of this from a variational standpoint is the initial QAOA procedure to discover the shape of $v(s)$.  Given the asymptotic nature of this curve, it is possible to find $v(s)$ for a given $p$ (corresponding to a QAOA procedure that takes time $t_f$) and then scale it up into a bang-anneal-bang ansatz for a larger $t_f$.

Since the base annealing curve is related asymptotically to an optimized adiabatic schedule, it could be possible to use insight from the adiabatic limit to bypass the QAOA step entirely and create an ansatz for $v(s)$ \textit{a priori}.  For instance, in the unstructured search problem, it could be possible to use Roland \& Cerf's \cite{Roland} optimized adiabatic annealing schedule as a guess for the $u_0(t)$ base curve.  If knowledge of the spectral gap is available, similar curves could be constructed for other problem instances.

\begin{algorithm}
\caption{Bang-Anneal-Bang Ansatz Optimization}
\label{alg:bab}
\begin{algorithmic}[1]
\Procedure{BAB Ansatz Algorihm}{p}
    \State $\beta_i,\, \gamma_i \gets \textrm{OptimizeQAOA}(p)$ 
        \Comment{Find QAOA angles}
    \State $v_0(\frac{i}{p-1}) \gets \frac{\beta_i}{\beta_i+\gamma_i}$ 
        \Comment{Approximate annealing curve}
    \State $T\gets\sum_{i=0}^{p-1} |\beta_i|+|\gamma_i|$ 
        \Comment{Total QAOA time}
    \State $\tilde\beta,~\tilde\gamma,~\omega,~\phi \gets\text{Initial Guess}$
    
    \While{Optimizing}
        \State $v_{bab}(t) \gets \textsc{ConstructAnsatz}(v_0(s), \tilde{\beta},\tilde{\gamma},T,\omega,\phi)$
        
        \State $E_{bab} \gets \textrm{Evolution under~} v_{bab}$
            \Comment Energy of Protocol
        \State $\tilde\beta,~\tilde\gamma,~\omega,~\phi \gets \text{Update Based on Optimization}$
        
    \EndWhile
    \State $v_{bab}(t) \gets \textsc{ConstructAnsatz}(v_0(s), \tilde{\beta},\tilde{\gamma},T,\omega,\phi)$
        \Comment{Final Protocol}
    \State \textbf{return} $v_{bab}(t)$
      
\EndProcedure

\Function{ConstructAnsatz}{$v_0(s), \tilde{\beta},\tilde{\gamma},T,\omega,\phi$}

    \For{$t\in [0,\tilde\beta]$} \Comment{Initial Bang}
        \State $v_{bab}(t) \gets 0$
    \EndFor
    \For{$t_k\in [\tilde\beta, T-\tilde\gamma]$} \Comment{Interior Anneal}
        \State $v_{bab}(t) \gets v_0(\frac{t-\tilde{\gamma}}{T-\tilde{\gamma}-\tilde{\beta}}) + A \cos{(\omega\, t+\phi)}$
    \EndFor
    \For{$t_k\in [T-\tilde\gamma,T]$}\Comment{Final Bang}
        \State $v_{bab}(t_k) \gets 1$
    \EndFor
\State \textbf{return} $v_{bab}(t)$
\EndFunction
\end{algorithmic}
\end{algorithm}

Below we present some of the results for this algorithm, simulated on a classical computer, solving directly the Schr\"odinger equation.  Our algorithm could be implemented on a quantum computer, replacing this simulation of the Schr\"odinger equation with actual quantum evolution.  Three different levels of the above algorithm are used.  The first listed as ``Basic Interpolation'' is the form used by \cite{Zhou2018} where the QAOA derived asymptotic curve $v(s)$ is just interpreted as an annealing curve.  The second, ``Sine Interpolation,'' superposes on top of this a sine curve whose period is equal to the average duration of a QAOA layer.  Finally, ``BAB Ansatz'' is our full Bang-Anneal-Bang Ansatz, here treating $\tilde{\gamma}$ (the initial bang), $\tilde{\beta}$ (the final bang), $\omega$ (the frequency of the oscillations), $A$ (the amplitude of the oscillations), and $\phi$ (the phase of the oscillations) to all be variational parameters.  This version of the algorithm is outlined in pseudocode in Alg.~\ref{alg:bab}.  For all procedures, the time allotted is $t_f$, the same as the time that the QAOA procedure took.

\begin{figure}
        \centering
                \includegraphics[width=0.5\textwidth]{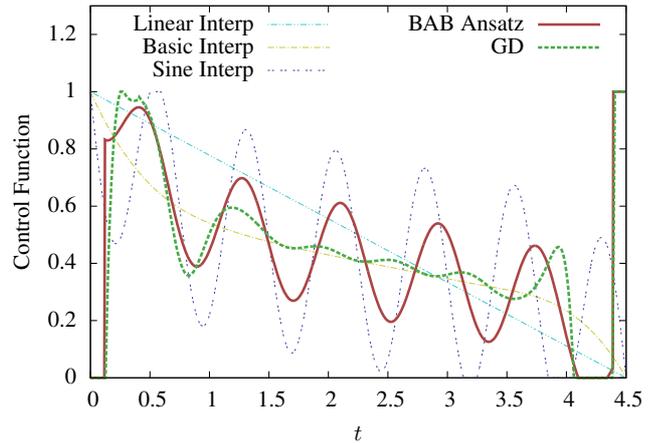}
        \caption{Various versions of our algorithm, with ``BAB Ansatz'' being the most advanced in terms of number of parameters.  The BAB Ansatz approximately follows the optimal procedure found using gradient descent, ``GD.''  The energies resulting from each procedure can be found in Tab.~\ref{tab:bab_ansatz_example_1}.  The problem instance in this case is the same as in Fig.~\ref{fig:qaoa_curves} but is run for a time corresponding to $p=6$ QAOA. }
        \label{fig:bab_ansatz_example_1}
\end{figure}

Fig.~\ref{fig:bab_ansatz_example_1} shows an example.  The kickstarting QAOA procedure used $p = 6$, and also shown are a basic linear ramp, and the exact optimal procedure found via a gradient descent (``GD'') procedure.  The resulting energies of all the relevant procedures are summarized in the accompanying  Table.~\ref{tab:bab_ansatz_example_1}.  While this represents results for a single problem instance, these results are typical of what we see in other instances.  Notably, the relative ordering of the energies achieved by each procedure in Table~\ref{tab:bab_ansatz_example_1} is typical of all instances we examined.
\begin{table}
        \centering
                \begin{tabular}{c|c}
                Model&Energy\\\hline
                Linear& -5.987\\
                Basic&-6.257\\
                Sine&-6.522\\
                QAOA&-6.578\\
                BAB&-6.636\\
                GD&-6.705\\
                Ground&-7.214
                \end{tabular}
        \caption{Energies related to the procedures shown in Fig.~\ref{fig:bab_ansatz_example_1}.  The more information is used in constructing the ansatz, the closer we get to the optimal energy possible for this time.  Notably, our full BAB Ansatz is required to perform better than the underlying QAOA protocols used to kickstart this procedure.  Based on our numerics, this ordering is representative of the relative qualities of the algorithms with BAB outperforming QAOA but not quite reaching the quality of the Optimal procedure.}
        \label{tab:bab_ansatz_example_1}
\end{table}

Note that we have attempted to run this algorithm without the initial QAOA procedure to find an estimate for $v(s)$ and instead using just a linear ramp for $v(s)$.  The resulting procedure performs poorly compared to any of our procedures employing the QAOA derived asymptotic curve but is still favorable when compared to a simple linear ramp.  As stated before, if anything is known \textit{a priori} about the shape of this base annealing curve, that information can be used instead of the QAOA procedure.

\section{Conclusion}
\label{sec:conc}

The optimal protocol is by construction the most efficient way to operate a quantum annealer or analog quantum computer.  This protocol demonstrates structure including the initial and final bangs explored in Ref.~\cite{Brady2020} that vanish in the long-$t_f$ limit.

This work explored the structure of the annealing region in more detail.  Due to previous results regarding the optimality of the adiabatic path \cite{Bukov}, we expect and indeed see that in the long-$t_f$ limit, the annealing region approaches an optimized adiabatic schedule, similar to what was derived by Roland and Cerf \cite{Roland}.  Furthermore, that optimal curve's annealing section has an oscillatory pattern superposed on top of it.  In the adiabatic limit, the amplitude of these oscillations should vanish to recover a monotonic annealing ramp.  However, in the near-adiabatic limit, these oscillations are helpful in managing the leakage between the ground state and first excited state.  We derive the near-adiabatic form of these oscillations and describe their dependence on the phase difference between the ground state and first excited state amplitudes outside of this perturbative limit.

This analysis of the near-adiabatic limit of the annealing curve should be of interest in itself since it can be used to potentially enhance adiabatic protocols with little additional \textit{a priori} information.  This enhancement can be implemented by our algorithm in Sec.~\ref{sec:bab_ansatz} using the original annealing curve instead of a QAOA-derived curve.

Furthermore, we explore the connections between QAOA and this oscillatory structure of the optimal curves.  Numerically, we see that optimal QAOA schedule incorporates the structure of  the underlying optimal curve.  The length of the QAOA layers matches up with the oscillation period of the annealing curve, and the ratio of the bang lengths within the QAOA layer matches up with the average value of the annealing curve within that period.

This behavior provides an explanation for the QAOA asymptotic curve behavior at large $p$ seen in Refs.~\cite{Zhou2018, Pagano2019}.  The behavior of the optimal curve can be understood asymptotically where it approaches an optimized adiabatic procedure with a fixed curve form.  If QAOA is emulating this optimal curve, then QAOA should also be approaching a fixed asymptotic form.

We sought to provide analytic evidence for this matching up between the QAOA curve and the optimal procedure.  Our results do show that there is an decrease in the error of a product formula if the product formula step size matches the oscillations in an annealing curve being Trotterized.  Furthermore, this enhancement requires that the ratio of the bangs follows the annealing curve, just as we see in the numerics.  Unfortunately, this error bound scales unfavorably with $p$, the number of QAOA layers, failing to match up with the scaling in practice.  Based on other methods, we provide further arguments for how this additional scaling behavior could occur, but it remains an open question how to tighten this analysis to match the exact scaling seen in QAOA in practice.

As a result of this analytic and numeric work, we not only achieve an explanation for the asymptotic large-$p$ behavior of QAOA, but we also better understand the optimal procedure.  One of the main difficulties with the optimal curve is that it is not feasible to construct this protocol on real hardware.  The protocol requires too much information about the intermediate quantum state and requires treating an entire smooth curve as a variational parameter.  To address these issues, in Section \ref{sec:bab_ansatz}, we constructed a new algorithm that uses the results of this paper to create an ansatz with very few variational parameters that outperforms naive quantum annealing and QAOA. 

This algorithm uses a QAOA procedure to find the form of the annealing region of the optimal procedure and then uses this to create an ansatz.  This ansatz then treats the lengths of the initial bang, final bang, and some basic properties of the oscillatory pattern as variational parameters in an ansatz.  In practice, this algorithm outperforms QAOA and quantum annealing but falls slightly short of the full optimal protocol.

\acknowledgments

We would like to thank Chris Baldwin and Minh Tran for helpful discussions.

The research of L.T.B. was partially supported by a National Institute of Standards and Technology (NIST) National Research Council (NRC) Research Postdoctoral Associateship Award in the Information Technology Lab (ITL).
The research was supported by the 
U.S. Department of Energy Award No.\ DE-SC0019449 for work analyzing the structure of QAOA curves,
AFOSR for work analyzing and exploring the perturbative near-adiabatic limit,
and ARO MURI for work associated with the asymptotic behavior and continuous limit of QAOA.

This material is based upon work supported by the U.S. Department of Energy, Office of Science, Office of Advanced Scientific Computing Research, under the Quantum Computing Application Teams program.  Sandia National Laboratories is a multimission laboratory managed and operated by National Technology \& Engineering Solutions of Sandia, LLC, a wholly owned subsidiary of Honeywell International Inc., for the U.S. Department of Energy’s National Nuclear Security Administration under contract DE-NA0003525. This paper describes objective technical results and analysis. Any subjective views or opinions that might be expressed in the paper do not necessarily represent the views of the U.S. Department of Energy or the United States Government.

\appendix

\section{Near-Adiabatic}
\label{app:NA}
In this section of the appendix, we explore additional features of the near-adiabatic limit.  In section \ref{app:NA_pert}, we carry out a perturbative analysis of the near-adiabatic equations to find the form of the oscillations, $c(t)$, in the limit where the base ramp is changing slowly $\dot{u}_0\ll1$.  The end result of this section is the derivation of Eq.~(\ref{eq:c_perturbative_limit}) from the main text.

The derivation of the near-adiabtic limit in the main paper relies on following the base annealing curve without the oscillations.  It is possible to derive the near-adiabatic frame following the exact adiabatic frame, following the control function, oscillations and all.  We do this derivation in Section \ref{app:NA_AF}.  This formulation is more useful outside the perturbative limit.  This connects to Eq.~(\ref{eq:Theta2}) from the main text.

Section \ref{app:NA_opt} derives the optimal control equations in the near-adiabatic limit.  These differential equations, if solvable, would give not only the oscillatory portion $c(t)$ but the entire function $u(t) = u_0(t)+c(t)$, enforcing $u(0)$ and $u(t_f)$.  These equations might be of interest to experts or numericists but no longer lend themselves to a perturbative analysis, making them less useful within the current context.

\subsection{Perturbative Limit}
\label{app:NA_pert}

Recall the Schr\"odinger equation, Eq.~\eqref{hA} and \eqref{hB}, for the ground state and first excited state amplitudes,
\begin{align}
    &i\left(\der{C_0(t)}{t} + C_1(t) \frac{\gamma(t)\dot{u}_0(t)}{\Delta(t)} \right)\label{hA2}\\\nonumber
    &= c(t)\left(C_0(t)\kappa_0(t)+C_1(t)\gamma(t)\right),
\end{align}
\begin{align}
    &i\left(\der{C_1(t)}{t} - C_0(t)\frac{\gamma(t)\dot{u}_0(t)}{\Delta(t)}\right)\label{hB2}\\\nonumber
    &=\Delta(t)C_1(t)+c(t)\left(C_0(t)\gamma(t)+C_1(t)\kappa_1(t)\right).
\end{align}

Next, we will Taylor expand $\Delta(u_0(t))$ and $\gamma(u_0(t))$ around $u_0(0) = u^{(0)}_0$ so that
\begin{align}
    \label{eq:expansion_Delta}
    \Delta(u_0(t)) &\approx \Delta(u_0^{(0)})+\left.\der{\Delta(u_0(t))}{u_0(t)}\right|_{u_0(t)\to u_0^{(0)}}\dot{u}_0 t,\\
    \gamma(u_0(t)) &\approx \gamma(u_0^{(0)})+\left.\der{\gamma(u_0(t))}{u_0(t)}\right|_{u_0(t)\to u_0^{(0)}}\dot{u}_0 t.
\end{align}
For convenience, we shorten the notation here so that 
\begin{align}
    \Delta(u_0(t)) &\approx \Delta_0+\Delta_1\dot{u}_0 t,\\
    \gamma(u_0(t)) &\approx \gamma_0+\gamma_1\dot{u}_0 t.
    \label{eq:expansion_gamma_2}
\end{align}

Officially, we would need to do the same expansion with the $\kappa$ variables, but it turns out that the $\kappa_j$ are not relevant to lowest non-zero order in perturbation theory.

The strategy here will be to do time dependent perturbation theory, specifically keeping track of orders of $\dot{u}_0$.  Furthermore, we will assume that we are zoomed in on one section of the $u_0(t)$ curve where $\dot{u}_0$ can be treated as approximately constant.

We will furthermore make an ansatz that our additional control function 
\begin{equation}
    \label{eq:app_ansatz}
    c(t)=c_0\sin( \omega t+\theta).
\end{equation}

In the new notation, we have two-level Hamiltonian that we will
 analyse within time-dependent perturbation theory with 
\begin{eqnarray}
H_0=\left(
\begin{array}{cc}
 0 & 0 \\
 0 & \Delta _0 \\
\end{array}
\right),
\end{eqnarray}
and perturbation
\begin{widetext}
\begin{equation}
\label{eq:V_matrix}
V= \left(
\begin{array}{cc}
 c(t) \kappa _0 & (\gamma_0+\gamma_1 \dot{u}_0 t)\left(c(t)  -i \frac{\dot{u}_0}{\Delta_0+\Delta_1 \dot{u}_0 t}\right)  \\
 (\gamma _0+\gamma_1 \dot{u}_0 t)\left(c(t)  +i\frac{\dot{u}_0}{\Delta_0+\Delta_1 \dot{u}_0 t}\right) & c(t) \kappa _1+\Delta _1\dot{u}_0 t \\
\end{array}
\right).
\end{equation}
\end{widetext}

If we assumed that $c_0\in\mathcal{O}(\dot{u}_0)$ and did a perturbative expansion to first order in $\dot{u}_0$, we would find that $c_0=0$ leads to no leakage between the ground state and first excited state.  Therefore, in order to get a nontrivial solution, we assume that $c_0\in\mathcal{O}(\dot{u}_0^2)$.
Furthermore, looking at $V$, we can see that it is $V\in\mathcal{O}(\dot{u}_0)$.  Therefore, if we look at time dependent perturbation theory to second order in $V$, we will extract all the second order dependence on $\dot{u}_0$.
Up to the second order in the perturbation $\mathcal{V}$, 
\begin{eqnarray}
\ket{\psi_I(t)} = \Big{[} 1-i \int_0^t d\tau\! \mathcal{V}(t) \!
 -\! \int_0^t d\tau\! \mathcal{V}(\tau)\! \int_0^\tau d\tau_2\!  \mathcal{V}(\tau_2) \nonumber
 \Big{]} \ket{\psi(0)},
\end{eqnarray}
with $\mathcal{V}(t)=W(t)^\dagger V(t) W(t)$ where 
\begin{eqnarray}
\label{eq:app_W}
W(t)=\left(
\begin{array}{cc}
 1 & 0 \\
 0 & e^{- i\Delta _0 t} \\
\end{array}
\right).
\end{eqnarray}
In order to  transform the wave-function from the interaction picture $\psi_I\to \psi$ back to original representation, we use relation $\ket{\psi(t)}=W(t)\ket{\psi_I(t)}$.

We can perform these integrals keeping terms  up to second order in $\dot{u}_0$.  We specifically want there to be no leakage between the ground state and first excited state after one oscillation period.  The natural frequency in this system is $\Delta_0$, so we expect 
\begin{equation}
    \label{eq:app_omega}
    \omega = \Delta_0+\delta\omega
\end{equation}
and $t_f = \frac{2\pi}{\Delta_0}-\delta t_f$ where $\delta t_f,\delta \omega \in\mathcal{O}(\dot{u}_0)$.  Because $\omega$ only appears inside the trig function in Eq.~(\ref{eq:app_ansatz}) which is multiplied by $c_0\in\mathcal{O}(\dot{u}_0^2)$, $\delta\omega\in\mathcal{O}(\dot{u}_0)$ does not matter to the level of perturbation theory considered in this section.  To zeroth order in $\dot u_0$, $t_f=\frac{2\pi}{\Delta_0}$.  It is possible that $\omega$ could be farther away from $\Delta_0$, but numerical analysis of these equations as well as other evidence presented in the main body suggests that $\omega\approx \Delta_0$ leads to the smallest correction that still works.  Given the assumptions of the near-adiabatic approximations, a smaller correction term is preferable.

Let's define a unitary operator $U(t)$ via $
\ket{\psi_I(t)} = U(t) \ket{\psi(0)}
$, so that $\ket{\psi(t_f)} = W(t_f)U(t_f) \ket{\psi(0)}$.  We want no leakage between the ground state and the first excited state by the end of this evolution which means that we want $U_{12}(t_f)=0$ up to the relevant order.

After a lengthy calculation, one can show that
\begin{align}
    \label{eq:Upert}
    U_{12}(t_f) &= \frac{\gamma_0}{\Delta_0}\delta t_f \dot{u}_0-\frac{2\pi^2\gamma_0 \Delta_1}{\Delta_0^4}\dot{u}_0^2-\frac{\pi \gamma_0 \cos \theta}{\Delta_0}c_0\\\nonumber
    &+i\frac{4\pi\gamma_0 \Delta_1}{\Delta_0^4}\dot{u}_0^2-i\frac{2\pi\gamma_1}{\Delta_0^3}\dot{u}_0^2-i\frac{\pi\gamma_0\sin\theta}{\Delta_0}c_0+\mathcal{O}(\dot{u}_0^3).
\end{align}
Due to our assumptions about stoquasticity, all the quantities represented here are real which means that we can separate Eq.~\eqref{eq:Upert} 
out into real and imaginary components and require those two sets to sum to zero independently.  So we need the terms on the first line of Eq.~(\ref{eq:Upert}) to sum to zero independently and the terms on the second line to sum to zero independently.

Looking at the imaginary parts of Eq.~(\ref{eq:Upert}), we can see that the leakage is zero if
\begin{equation}
    c_0 = \dot{u}_0^2\frac{\csc\theta}{\Delta_0^2}\left(\frac{2\Delta_1}{\Delta_0}-\frac{\gamma_1}{\gamma_0}\right).
\end{equation}
The smallest amplitude of oscillations corresponds to $\theta = \frac{\pi}{2}$, that results in 
\begin{equation}
    c_0 = \frac{\dot{u}_0^2}{\Delta_0^2}\left(\frac{2\Delta_1}{\Delta_0}-\frac{\gamma_1}{\gamma_0}\right) + \mathcal{O}(\dot{u}_0^3).
    \label{eq:c0_first_order}
\end{equation}
By substituting $\theta\to\pi/2$ in Eq.~(\ref{eq:Upert}) and requiring the real part of the resulting expression to vanish we arrive at
\begin{equation}
    \label{eq:app_deltatf}
    \delta t_f = \frac{2\pi^2 \Delta_1}{\Delta_0^3}\dot{u}_0+\mathcal{O}(\dot{u}_0^2).
\end{equation}

So with this oscillation in the control function, we have proven that after a time $t_f = \frac{2\pi}{\Delta_0} - \delta t_f$, the amplitudes will return to themselves, resulting in perfect adiabatic transfer up to second order in $\dot{u}_0^2$ in this near-adiabatic limit.

We can now plug into our ansatz, Eq.~(\ref{eq:app_ansatz}), all the results of this section to get
\begin{equation}
    c(t) = \frac{\dot{u}_0(t)^2}{\Delta(u_0(t))^2}\der{\ln\left(\frac{\Delta(u_0(t))^2}{\gamma(u_0(t))}\right)}{u_0(t)}\cos(\Delta(u_0(t)) t) + \mathcal{O}(\dot{u}_0^3),
\end{equation}
which uses the results of Eq. (\ref{eq:c0_first_order}) as well as the definitions of $\Delta_1$ and $\gamma_1$ from Eqs. (\ref{eq:expansion_Delta}-\ref{eq:expansion_gamma_2}).  The log derivative is used to compress notation.  This form of oscillations will cancel out the deleterious effects of $\dot{u}_0\neq0$ up to second order.

The ansatz for $\omega$ in Eq.~\ref{eq:app_omega} is ultimately based off the zeroth order approximation of $\varphi(t)$ which determines the natural frequency in this problem.  The correction to the total evolution time  $\delta t_f$ could, therefore, be derived by considering  the phase difference between the first excited state and ground state.  We start with the $\dot{\varphi}$ equation, Eq.~(\ref{eq:na_na_phase}), and look for the solution to this equation to first order in $\dot{u}_0$ (which means that $c(t)$ will be negligibly small).  Firstly, since the $A_i$ terms only appear attached to small parameters, we can approximate them by their zeroth order constants, and the same goes for $\Delta$ and $\gamma$.  The $\sin\varphi$ and $\cos\varphi$ terms are troublesome, but they are already multiplied by small parameters, so we can approximate them by the zeroth order solution to this equation $\varphi(t) \approx \Delta^{(0)}t$ (using as a boundary condition that $\varphi(0)=0$).  With these approximations and iterations, the first order solution to the equation is
\begin{align}
    \varphi(t) &\approx \Delta_0t
    \\\nonumber
    &+\left(\frac{1}{2}\Delta_1 t^2+\frac{\gamma_0(\cos(\Delta_0t)-1)}{\Delta_0^2\tan(2\vartheta)}\right)\dot{u}_0.
\end{align}

We expect the populations to return after one full cycle of the system, both in terms of the phases of the eigenstate populations and the frequency of the ansatz $\omega$.
We can then ask what value of $t$ corresponds to a full period of oscillation such that $\varphi(t_f) = 2\pi+\mathcal{O}(\dot{u}_0^2)$.  To zeroth order in $\dot{u}_0$ it is obvious that $t_f = 2\pi/\Delta_0$.  Using this, it is easy to show that
\begin{equation}
    t_f = \frac{2\pi}{\Delta _0}- \frac{2\pi^2 \Delta_1}{\Delta_0^3}\dot{u}_0+\mathcal{O}(\dot{u}_0^2).
\end{equation}
corresponds to one period of oscillation for the phase.  Note that this time does indeed correspond to Eq.~(\ref{eq:app_deltatf}) derived through by requiring no leakage from the unitary matrix.

\subsection{Adiabatic Frame}
\label{app:NA_AF}
In the main text, we derived the near-adiabatic limit for the case of the instantaneous eigenframe evolving alongside the base curve $u_0(t)$ so that the full control function was given by $u(t) = u_0(t)+c(t)$.  In this setting $c(t)$ was our actual free function with $u_0(t)$ fixed and $c(0)=c(t_f)=0$.  It is possible to treat this entire problem a control problem and seek out the $u(t)$ that maintains populations the best between $t=0$ and $t=t_f$ subject to the constraint that $u(0) = u_a$ and $u(t_f) = u_b$.

The full version of this problem would just result in optimal curves in general, but here we are interested in just the smooth annealing region.  One feature of this smooth annealing region is that the bangs have already excited up some of the state into the first excited state, making the near-adiabatic approximation even more relevant.  For the purposes of this section, we will zoom in on one small region of the annealing curve and still implicitly assume that $u_a$ and $u_b$ are not that far apart.

We consider the probability amplitudes, $C_i(t)$ of being in the instantaneous eigenstates of a system, $\ket{j(u(t))}$, with instantaneous eigenenergies, $\lambda_j(u(t))$.  So our state can be written as
\begin{equation}
    \ket{\psi(t)} = \sum_{j} C_j(t)\ket{j(u(t))},
\end{equation}

Applying the Schr\"odinger equation to this state yields a set of coupled differential equations
\begin{align}
    &i\left(\der{C_k(t)}{t} + \sum_j C_j(t) \bra{k(u(t))}\der{}{t}\ket{j(u(t))} \right) 
    \\\nonumber
    &= \lambda_k(u(t)) C_k(t).
\end{align}

Now, a few assumptions will be made, the first being that the ground and first excited states are non-degenerate.  This could in general be satisfied by going into a symmetric subspace and looking at the relevant probability amplitudes within that symmetric subspace (for instance with the transverse field Ising model, we will consider the subspace defined by the usual Ising rotational symmetry).  The second and more relevant assumption is that $|C_0|\gg |C_1|\gg |C_2| \gg \ldots$ which is just a statement that we are in the near-adiabatic limit of evolution.  For our purposes, we will assume that the amplitudes for the second excited state and above are small enough throughout the evolution to be negligible.  Furthermore, we will later consider $|C_1|$ to be a small quantity relative to $|C_0|$ for approximation purposes.  The last requirement is that we set $\lambda_0(t) = 0$ which can be done without loss of generality.

Applying these assumptions and following the calculations of \cite{Brady2018}, we derive the equations:
\begin{align}
    \label{eq:C0}
    \der{C_0(t)}{t} &= -\frac{\gamma(u(t))\dot{u}(t)}{\Delta(u(t))}C_1(t),\\
    \label{eq:C1}
    \der{C_1(t)}{t} &= \frac{\gamma(u(t))\dot{u}(t)}{\Delta(u(t))}C_0(t)-i\Delta(u(t))C_1(t),
\end{align}
where the new functions represent
\begin{align}
    \gamma(u) &\equiv \bra{\varphi_0(u)}(\hat{B}-\hat{C})\ket{\varphi_1(u)},\\
    \Delta(u) &\equiv \lambda_1(u) - \lambda_0(u).
\end{align}

This leaves us with two coupled, complex differential equations.  To make things more explicit, we now split the $C$ variables into real amplitudes and phases such that
\begin{align*}
    C_0(t) &= e^{i\varphi_0(t)}A_0(t),\\
    C_1(t) &= e^{i\varphi_1(t)}A_1(t).
\end{align*}
These can be inserted into the differential equations.  After some algebra, including separating out real and imaginary components, the differential equations reduce to the real equations
\begin{align}
    \label{eq:phidot}
    \dot{\varphi} 
    &=
    \Delta-\frac{A_0^2-A_1^2}{A_0 A_1} \frac{\gamma \dot{u}}{\Delta}\sin(\varphi),
    \\
    \dot{A}_0 &= -\frac{\gamma \dot{u}}{\Delta}\cos(\varphi)A_1,
    \\
    \dot{A}_1 &= \frac{\gamma \dot{u}}{\Delta}\cos(\varphi)A_0,
\end{align}
where $\varphi(t) \equiv \varphi_0(t)-\varphi_1(t)$

The $A$ equations can be integrated to give
\begin{align}
    A_0(t) &= a\cos\left(
        \int_0^t dt'\frac{\gamma\dot{u}}{\Delta}\cos(\varphi)+\vartheta
    \right),\\
    A_1(t) &= a\sin\left(
        \int_0^t dt'\frac{\gamma\dot{u}}{\Delta}\cos(\varphi)+\vartheta
    \right).
\end{align}
These equations are similar to what was seen in the main text, and once again we are left with the conclusion that at the end of the evolution, we want 
\begin{equation}
    \Theta[u(t)] \equiv \int_0^{t_f} dt\frac{\gamma\dot{u}}{\Delta}\cos(\varphi)
\end{equation}
to be close to a multiple of $\pi$.  Though, we again have the caveat that having $\Theta[u(t)]$ equal to any multiple of $\pi$ other than zero would violate the assumptions of near-adiabaticity.

\subsection{Optimal Control}
\label{app:NA_opt}

Our setup is to take a procedure that goes from time $0$ to time $t_f$ moving from $u(0)=u_1$ at the beginning to $u(t_f)=u_2$ at the end.  We want to ensure that the instantaneous eigenstate populations are maintained during that evolution, at least from the beginning to the end (but not necessarily in the middle), so we want to minimize
\begin{align}
    J &= |C_0^*(t_f)C_0(t_f)-C_0^*(0)C_0(0)|\\\nonumber
    &+|C_1^*(t_f)C_1(t_f)-C_1^*(0)C_1(0)|.
\end{align}
The actual form of whether we are looking at the $L^1$ or $L^2$ norm of the difference between the probabilities is largely irrelevant, and another choice could be made with little consequence.  We have also written out the probabilities explicitly as $|C|^2 = C^* C$ which will be helpful shortly.

Now, we will treat this as an optimal control problem, seeking to find the conditions on $u(t)$ such that $J$ is minimized.  In order to enforce Eqs.~\ref{eq:C0}~\&~\ref{eq:C1}, we introduce Lagrange Multipliers $D_0(t)$ and $D_1(t)$ so that
\begin{align}
    J &= |C_0^*(t_f)C_0(t_f)-C_0^*(0)C_0(0)|\\\nonumber
    &+|C_1^*(t_f)C_1(t_f)-C_1^*(0)C_1(0)|\\\nonumber
    &+\int_{0}^{t_f}dt\ \bigg[ 
        D_0(t)\left(\dot{C}_0(t) +\frac{\gamma(u(t))\dot{u}(t)}{\Delta(u(t))}C_1(t)\right)\\\nonumber
    &+ D_1(t)\left(\dot{C}_1(t) - \frac{\gamma(u(t))\dot{u}(t)}{\Delta(u(t))}C_0(t)+i\Delta(u(t))C_1(t)\right)\bigg]\\\nonumber
    &+\cc
\end{align}
where the final $\cc$ indicates that we need to complex conjugates of the third and forth lines, just to treat the variables and their complex conjugates equally (remember that $u(t)$ is purely real).

Now, we just perform a Calculus of Variations analysis of this using $C_0(t)$, $C_1(t)$, $C^*_0(t)$, $C^*_1(t)$, and $u(t)$ as the variational parameters.  In doing this procedure, it is important to remember that the $C$ variables are fixed at $t=0$ but not at $t_f$ and that $u(t)$ is fixed at both end points.

Note that under a full optimal control theory analysis, such as \cite{Brady2020}, there would be restrictions on $u(t)$ such as $u(t)\in[0,1]$.  In this setting, we will ignore this restriction for ease of analysis, and this ignoring is justified by the fact that we are interested specifically at how this system behaves in an annealing region.  We are using this analysis explicitly to look at the annealing rather than bang-bang portions of the control function, and any of our results here should be taken explicitly within that context.  Also note that any restrictions on the $C$ variables is already taken care of by the fact that Eqs.~\ref{eq:C0}~\&~\ref{eq:C1} are being enforced by the Lagrange multipliers.  These equations came from the Schr\"odinger equation, so the $C$s will obey all necessary properties of probability amplitudes.

The resulting end point equations yield the boundary conditions for the $D$ variables 
\begin{align*}
    D_0(t_f)&=-\sgn(|C_0(t_f)|^2-|C_0(0)|^2)C_0^*(t_s), \\
    D_1(t_f)&=-\sgn(|C_1(t_f)|^2-|C_1(0)|^2)C_1^*(t_s).
\end{align*}
Any changes to using the $L^1$ or $L^2$ norm originally would have shown up here and would have just resulted in slightly different boundary conditions.

The variational procedure for the $D$ Lagrange multipliers just results in Eqs.~\ref{eq:C0}~\&~\ref{eq:C1} again as expected, and the variational procedure for the $C$ variables results in
\begin{align}
    \label{eq:D0}
    \dot{D}_0(t) &= - \frac{\gamma(u(t))\dot{u}(t)}{\Delta(u(t))}D_1(t),\\
    \label{eq:D1}
    \dot{D}_1(t) &= \frac{\gamma(u(t))\dot{u}(t)}{\Delta(u(t))}D_0(t) - i\Delta(u(t))D_1(t).
\end{align}
These are essentially following their own Schr\"odinger evolution.  Also note that based on the boundary conditions for the $D$s, we have $D_1(t)$ roughly the same size as $C_1(t)$ and $D_0(t)$ roughly the same size as $C_0(t)$.  Hence we can use the same hierarchy of $C_0\gg C_1$ with these new variables.

The last equation, resulting from the variations of $u(t)$ is the one that is actually important here.  Assuming the gap is nonzero, the resulting condition can be written as (suppressing functional dependencies for space reasons)
\begin{align}
    \gamma\left(
    D_0 \dot{C}_1+\dot{D}_0 C_1-C_0 \dot{D}_1 -\dot{C}_0 D_1
    \right)
    =
    i C_1 D_1\Delta \Delta',
\end{align}
where $\Delta' = \pder{\Delta(u(t)}{u(t)}$.  The natural next step is to use Eqs.~\ref{eq:C0},~\ref{eq:C1},~\ref{eq:D0},~\&~\ref{eq:D1} to eliminate the time derivatives of the $C$ and $D$ variables:
\begin{align}
    &\left(C_1(t) D_0(t)+D_0(t) C_1(t)\right) \gamma(u(t)) \\\nonumber
    =&- C_1(t) D_1(t) \Delta'(u(t)),
\end{align}

This gives us the full set of optimal control equations that are necessary for an optimal procedure.  Unfortunately, this formalism does not lend itself to the perturbative analysis discussed in the previous sections.  These results are presented for completeness.

\section{Trotterization Error}
\label{app:trot_error}

In this appendix we examine product formula errors and how they interact with the oscillations observed in the optimal curve.

In Sections \ref{app:trot_error_op} and \ref{app:trot_error_adi} we derive directly how these oscillations influence the product formula.  Section \ref{app:trot_error_op} focuses on the standard product formula error formulated in terms of operators while Section \ref{app:trot_error_adi} follows the arguments of Ref.~\cite{Yi2021} and examines the product formula error for an adiabatic evolution.  In section \ref{app:trot_error_op} we derive Eq.~(\ref{eq:Trotter_Product_final}) from the main text.

The last two sections provide additional background for Section \ref{app:trot_error_op} with Section \ref{app:trottererror} rederiving the basic known formulas necessary to bound the Product Formula errors.  Section \ref{app:PF_perturbations} looks at the robustness of our oscillatory enhancement to perturbations of the product formula parameters.

\subsection{Standard Operator Error Scaling}
\label{app:trot_error_op}

In this section, our goal will be to find the bound on the matrix norm error between unitaries given in Eqs. \ref{eq:unitary_timeevo}~\&~\ref{eq:unitary_product} under the assumption that the evolution is governed by the oscillatory function given in Eq.~\ref{eq:utwsine}

This sequence of arguments will initially follow the appendix of \cite{Huyghebaert}.  For another good reference on this, try \cite{Poulin}.

Finding the error between these two unitaries is fairly straightforward and is laid out well in Ref.~\cite{Huyghebaert}.  We rederive this result in Appendix \ref{app:trottererror}.  To cite the result
\begin{align}
    &||\hat{U}(t_f,0)-\hat{U}_{PF}(t_f,0)||
    \\\nonumber
    &\leq\sum_{k=0}^{p-1} \int_{k\Delta t}^{(k+1)\Delta t} ds \int_{k \Delta t}^s dr \left|\left|\comm{\hat{H}_0(r)}{\hat{H}_1(s)}\right|\right|.
\end{align}
The matrix norm used in this proof was the standard operator norm.  Also, notably, this result does not rely on perturbative methods like the Baker-Campbell-Hausdorff equation or the Magnus expansion.

Now, it is fairly straightforward to specify down to the form we are using in which case
\begin{align}
    \label{eq:trot_error_1}
    &||\hat{U}(t_f,0)-\hat{U}_{PF}(t_f,0)||
    \\\nonumber
    &\leq
    \left|\left|\comm{\hat{B}}{\hat{C}}\right|\right|
    \sum_{k=0}^{p-1} 
    \int_{k\Delta t}^{(k+1)\Delta t} ds \int_{k \Delta t}^s dr\,
    u(r)(1-u(s)).
\end{align}

Next, the form in Eq.~(\ref{eq:trot_error_1}) is a little unruly to work with.  It actually is much easier to go over to Fourier space where
\begin{equation}
    u(t) = \int_{-\infty}^{\infty} d\xi\, \tilde{u}(\xi) e^{2\pi i t\xi}.
\end{equation}
Of course if we take the Fourier transform of Eq.~(\ref{eq:utwsine}), we would get
\begin{equation}
    \label{eq:uxi}
    \tilde{u}(\xi) = \tilde{u}_0(\xi) + \frac{c_0}{2i}
    \left(
        e^{i\phi}
        \delta\left(\xi+\frac{1}{\tau}\right)
        -
        e^{-i\phi}
        \delta\left(\xi-\frac{1}{\tau}\right)
    \right).
\end{equation}
Putting this Fourier transformed version in allows us to easily do the integrals over $s$ and $r$, resulting in
\begin{align}
    &||\hat{U}(t_f,0)-\hat{U}_{PF}(t_f,0)||
    \\\nonumber
    &\leq
    \left|\left|\comm{\hat{B}}{\hat{C}}\right|\right|
    \sum_{k=0}^{p-1} 
    \int_{-\infty}^{\infty}d\xi
    \int_{-\infty}^{\infty}d\eta\,
    \tilde{u}(\xi)\left(\delta(\eta)-\tilde{u}(\eta)\right)
    \\\nonumber
    &\times
    \left[\frac{\left( e^{2 i \pi  \Delta t \eta } \left(1-e^{2 i \pi  \Delta t \xi }\right)\eta-  \left(1-e^{2 i \pi  \Delta t \eta }\right)\xi\right) e^{2 i \pi  \Delta t k (\eta +\xi )}}{4 \pi ^2 \eta  \xi  (\eta +\xi )}\right].
\end{align}
Notice that all the $k$ dependence is in the last line, so we can carry out the $k$ sum fully to get
\begin{align}
    &||\hat{U}(t_f,0)-\hat{U}_{PF}(t_f,0)||
    \\\nonumber
    &\leq
    \left|\left|\comm{\hat{B}}{\hat{C}}\right|\right|
    \int_{-\infty}^{\infty}d\xi
    \int_{-\infty}^{\infty}d\eta\,
    \tilde{u}(\xi)\left(\delta(\eta)-\tilde{u}(\eta)\right)
    \\\nonumber
    &\times
    \frac{\left(\eta  \left(-1+e^{2 i \pi  \Delta t \xi }\right)+\xi  \left(1-e^{2 i \pi  \Delta t \eta }\right)\right) \left(1-e^{2 i \pi  \Delta t p (\eta +\xi )}\right)}{4 \pi ^2 \eta  \xi  (\eta +\xi ) \left(e^{2 i \pi  \Delta t \xi }-e^{-2 i \pi  \Delta t \eta }\right)}.
\end{align}

From this point, putting in Eq.~(\ref{eq:uxi}) and simplifying down is quite possible; although, the fully general expression is a bit messy and not horribly informative.
One possible simplification that is quite informative is the case where $\tau\to\Delta t$ in which case manipulation can simplify all of this nicely down to
\begin{align}
    &||\hat{U}(t_f,0)-\hat{U}_{PF}(t_f,0)||
    \\\nonumber
    &\leq
    \left|\left|\comm{\hat{B}}{\hat{C}}\right|\right|
    \int_{-\infty}^{\infty}d\xi
    \int_{-\infty}^{\infty}d\eta\,
    \tilde{u}_0(\xi)\left(\delta(\eta)-\tilde{u}_0(\eta)\right)
    \\\nonumber
    &\times
    \frac{\left(\eta  \left(-1+e^{2 i \pi  \Delta t \xi }\right)+\xi  \left(1-e^{2 i \pi  \Delta t \eta }\right)\right) \left(1-e^{2 i \pi  \Delta t p (\eta +\xi )}\right)}{4 \pi ^2 \eta  \xi  (\eta +\xi ) \left(e^{2 i \pi  \Delta t \xi }-e^{-2 i \pi  \Delta t \eta }\right)}
    \\\nonumber
    &-\left|\left|\comm{\hat{B}}{\hat{C}}\right|\right|
    c_0\frac{\Delta t^2 p \cos(\phi)}{2 \pi }.
\end{align}
If we undo the Fourier transform, this further reduces to
\begin{align}
    \label{eq:trot_error_final}
    &||\hat{U}(t_f,0)-\hat{U}_{PF}(t_f,0)||
    \\\nonumber
    &\leq
    \left|\left|\comm{\hat{B}}{\hat{C}}\right|\right|
    \sum_{k=0}^{p-1} 
    \int_{k\Delta t}^{(k+1)\Delta t} ds \int_{k \Delta t}^s dr\,
    u_0(r)(1-u_0(s))
    \\\nonumber
    &-\left|\left|\comm{\hat{B}}{\hat{C}}\right|\right|
    c_0\frac{\Delta t^2 p \cos(\phi)}{2 \pi }.
\end{align}
In other words, the sine function we added onto the control function essentially becomes decoupled from the rest of the error in the case that its period is the same as the Trotter slice size.  Furthermore, the first line of the error bound will always be positive (remember that $u_0\in[0,1]$), but the second line can be negative, effectively reducing the error in the Trotterization.  As promised, choosing $\phi=0$ results in the maximum improvement.

The improvement in the error is proportional to $\Delta t^2 p$ which is coincidentally the same rough scaling as the term above, so this term will actually be competitive and could contribute greatly to the error bound.  To see this more precisely, note that the first term in the bound can be upper bounded quite easily by $\frac{1}{2}\Delta t^2 p$ so that
\begin{align}
    &||\hat{U}(t_f,0)-\hat{U}_{PF}(t_f,0)||
    \\\nonumber
    &\leq
    \left|\left|\comm{\hat{B}}{\hat{C}}\right|\right|
    \frac{\Delta t^2 p}{2}
    \left(1-\frac{c_0}{\pi}\cos(\phi)\right).
\end{align}
Note that $c_0\leq0.5$ at the very worst to ensure that $u(t)\in[0,1]$, so it is not possible for this bound to be below zero.

In Appendix \ref{app:PF_perturbations} we explore the robustness of this effect to perturbations.

\subsection{Adiabatic Trotter Error}
\label{app:trot_error_adi}

In this subsection our goal will be to bound the Trotter error by looking at the error on the ground state fidelity directly.  It should be noted that our analysis indicates that the underlying annealing curve adiabatically transfers not just the ground state but also higher excited states, with this reducing down to just ground state adiabaticity in the limit of $t_f\to\infty$.  The results in this section focus on just the ground state, but similar results can be derived for any excited states, and those results can be simultaneously applicable.

The methods in this section closely follow the results of Yi and Crosson \cite{Yi2021} who themselves draw inspiration from \cite{Jansen} and \cite{Tran2020}.  Specifically, this result can be thought of as a modification of their Proposition 1 (proven in their Appendix F) to the setting where the underlying annealing curve has an oscillatory structure.  In practice, this modification is exactly the same as the modification to the usual Trotter operator error formula, meaning we can recover the oscillatory enhancement and still have the improved scaling analysis of Yi and Crosson.

As a reminder, the control function is
\begin{equation}
    u(t) = u_0(t/t_f) + c(t,t_f),
\end{equation}
where $u_0(s)$ is a smooth monotonically decreasing function, and
\begin{equation}
    c(t,t_f) = - c_0(t_f) \sin\left(\frac{2\pi}{\tau} t + \phi\right).
\end{equation}

We discretize our adiabatic evolution over time \(t_f\) into \(M\) steps with ``short'' timestep \(\Delta t = t_f/M\):
\begin{align}
    &\hat{U}_1\left(t+\Delta t, t\right)\nonumber\\
    =& \exp\left(-i \int^{t+\Delta t}_t\,\mbox d t' (u(t')\hat B+(1-u(t'))\hat C)\right).
\end{align}
We want to evaluate the integral
\begin{equation}
    \int^{t+\Delta t}_t \mbox d t' u(t') = \int^{t+\Delta t}_t \mbox d t' \left(u_0(\frac{t}{t_f})+c(t,t_f)\right).
\end{equation}
For our purposes, we will assume that $u_0(t/t_f)$ is approximately constant over this interval, which is valid for large $t_f$.  In other words, we assumed here that $u_0'(\frac{t}{t_f})$ is approximately constant over this small interval.  We will also now introduce $s\equiv \frac{t}{t_f}$ as a normalized time.  In this case
\begin{align}
    &\int^{t+\Delta t}_t \mbox d t' u(t') \\\nonumber&= \Delta t u_0(s) 
    + c_0 \frac{\tau}{2 \pi} (\cos(\frac{2 \pi}{\tau} (t+\Delta t) + \phi) - \cos (\frac{2 \pi}{\tau} t + \phi) ).
\end{align}
We can use trig identities to reduce this further to
\begin{align}
    \int^{t+\Delta t}_t \mbox d t' u(t') &= \Delta t u_0(s) 
    \\\nonumber
    &- c_0 \frac{\tau}{\pi} \sin(\frac{\pi\Delta t}{\tau})\sin(\frac{2 \pi}{\tau} (t+\frac{\Delta t}{2}) + \phi).
\end{align}
We will now define $\sigma\equiv \tau/t_f$ and $\Delta s \equiv \Delta t/t_f$ so that
\begin{align}
    \int^{t+\Delta t}_t \mbox d t' u(t') &= \Delta t U(s).
\end{align}
where for convenience, we have defined
\begin{equation}
    U(s)\equiv u_0(s) 
    - c_0 \frac{\sigma}{\Delta s \pi} \sin\left(\frac{\pi\Delta s}{\sigma}\right)\sin\left(\frac{2 \pi}{\sigma} (s+\frac{\Delta s}{2}) + \phi\right).
\end{equation}
In the limit of $\Delta s\to 0$, this just reduces to $U(s)\to u(s)$.

To recover the original results of \cite{Yi2021}, take $u_0=1-s$ for a linear ramp and $c_0=0$ for no oscillations. This is part of the discretization error that we will later assume is smaller than the adiabatic error.  

It is also important to note that $c_0\ll u_0'(s)$ (specifically elsewhere we found that $c_0\in\mathcal{O}(\dot{u}_0^2)$).  This means that the sinusoids can be counted as a correction to the $u_0(s)$ terms rather than their own term.  This will be useful when bounding quantities since we can then treat these two as a whole rather than seperate quantities to bound

Each such discrete unitary is then Trotterized to first order:
\begin{align}
    \hat{U'}_1\left(t+\Delta t, t\right) =& \exp\left(-i \Delta t U(s) \hat{B}\right)
    \\\nonumber
    &\times
    \exp\left(-i\Delta t (1-U(s))\hat{C} \right),
\end{align}
we define the effective Hamiltonian for this Trotterized evolution by
\begin{equation}
        \tH(t) = i\log\left(\hat{U'}_1\left(t+\Delta t, t\right)\right)/\Delta t.
\end{equation}
In the limit of the discretization step size $\Delta t\to 0$, there is a continuous Hamiltonian defined by this. This effective Hamiltonian has the nice property that $\tH(0) = \hat{B}$ and $\tH(t_f) = \hat{C}$, so evolution under $\tH(t)$ for slow $t_f$ can be described as an adiabatic process.  The optimal curves approach an adiabatic procedure with vanishing initial and final bangs in the large $t_f$ limit, so this is appropriate in our setting for large $t_f$ (corresponding to large $p$ for QAOA).

The core idea of this method then is to bound the error on the evolution, not using operator errors but using the adiabatic theorem directly.  This will result in tighter scaling in terms of the number of Trotter or QAOA slices $p$ but will introduce scaling with the $p$-independent spectral gap of $\tH(t)$.  This technique will use the adiabatic theorem of Jansen, Ruskai, and Seiler \cite{Jansen}.

We split the effective Hamiltonian into two parts such that
\begin{align}
        e^{-i \Delta t \tH(s)}  &= \left(e^{-i \Delta t U(s) \hat{B}}e^{i \Delta t U(s) \hat{C}}\right)e^{-i\Delta t \hat{C}}\\
        &\equiv e^{-i \Delta t \hat{G}(U(s))}e^{-i\Delta t \hat{C}}.
\end{align}  

To reiterate our ultimate goal, we would want to show that there is an enhancement when $\Delta t = \tau$.  As we will see, this goal is not consistent with the assumptions of this method, which we will discuss later in this section.

In order to utilize the Adiabatic condition bounds of \cite{Jansen}, it is necessary to compute the matrix norms of derivatives of $\tH(s)$ with respect to $s$.
Using Magnus expansion techniques, \cite{Yi2021} bounds the norms of these derivatives by
\begin{align}
        \label{eq:bound_tHp}
        ||\der{}{s}\tH||\leq& ||\der{}{s}\hat{G}||~ \mathcal{F}_1(2\Delta t||\hat{C}||+2\Delta t||\hat{G}||),\\
        ||\dder{}{s}\tH||\leq& ||\dder{}{s} \hat{G}||~\mathcal{F}_1(2\Delta t||\hat{C}||+2\Delta t||\hat{G}||)\\\nonumber
                &+2\Delta t||\der{}{s}\hat{G}||^2~ \mathcal{F}_0(2\Delta t||\hat{C}||+2\Delta t||\hat{G}||).
\end{align}
Here the $\mathcal{F}$ functions are defined to be
\begin{align}
        \mathcal{F}_0(x) &= \sum_{j=0}^\infty x^j = \frac{1}{1-x},\\
        \mathcal{F}_1(x) &= \sum_{j=1}^\infty \frac{x^{j-1}}{j} = -\ln(1-x)/x,\\
        \mathcal{F}_2(x) &= \sum_{j=2}^\infty \frac{x^{j}}{j^2} = -\int_0^x dx'\ln(1-x').
\end{align}
We present both norm derivatives here, but since our results only effect the constant prefactors and not overall scaling, it will be enough to keep track of $||\der{}{s}\tH||$ because our results do not alter the dominant term in Ref.~\cite{Yi2021} which depends on $||\der{}{s}\tH||$.

The next step in this process is to bound the norms of the derivatives of $\hat{G}$.  This is possible by looking at

\begin{equation}
        \label{eq:Magnus_DE}
        \der{}{s} e^{-i\Delta t \hat{G}(U(s))} = \tilde{E}(s,\Delta t)e^{-i\Delta t \hat{G}(U(s))},
\end{equation}
where
\begin{equation}
        \tilde{E}(s,\Delta t) = i\Delta t\der{U}{s}\, e^{i\Delta t\,U(s) \hat{B}}(\hat{B}-\hat{C})e^{-i\Delta t\,U(s) \hat{C}}.
\end{equation}
The $\der{U}{s}$ is the only new portion of our results compared to Ref. \cite{Yi2021}, and to recover their results exactly, we would need to set $\der{U}{s}\to 1$.  In our setting, this derivative is just
\begin{align}
        \der{U}{s} =  u_0'(s)- \frac{2c_0}{\Delta s} \cos\left(\frac{2\pi}{\sigma}(s+\frac{\Delta s}{2})+\phi\right)\sin\left(\frac{\pi\Delta s}{\sigma}\right).
\end{align}
Again in the limit of small $\Delta s$, this reduces to just
\begin{align}
        \der{U}{s} \to \der{u}{s} =  u_0'(s)- \frac{2\pi c_0}{\sigma} \cos\left(\frac{2\pi}{\sigma}s+\phi\right).
\end{align}

A nice feature of Eq.~\ref{eq:Magnus_DE} is that it has an exact solution in the form of the Magnus Expansion.  The terms in the Magnus expansion can then be bounded as in Ref.~\cite{Yi2021}, and we follow a similar bounding but now keeping track of $\der{U}{s}$.
For instance, we can work out that the first two derivatives of $\hat{G}$ can have their norms bounded by
\begin{align}
        ||\hat{G}||&\leq U(s) D_-+\frac{1}{2\Delta t} \mathcal{F}_2\left(2\Delta t D_- U(s)\right),\\
        ||\hat{G}'||&\leq U'(s) D_-+\Delta t D_1 U(s)U'(s) \mathcal{F}_1\left(2\Delta t D_- U(s)\right).
\end{align}
Here we define $D_-\equiv||\hat{B}-\hat{C}||$, $D_0 \equiv ||\hat{C}||$, and $D_1\equiv ||\comm{\hat{B}}{\hat{C}}||$.
Now, we can finally get back to the bound on $||\tH'||$ from Eq.~\ref{eq:bound_tHp}.
\begin{align}
        ||\tH'||&\leq ||\hat{G}'||~ \mathcal{F}_1(2\Delta t||\hat{C}||+2\Delta t||\hat{G}||)\\\nonumber
                &\leq \left[U'(s) D_-+\Delta t D_1 U(s)U'(s) \mathcal{F}_1\left(2\Delta t D_- U(s)\right)\right]\\\nonumber
                &~~~\times\mathcal{F}_1\left(2\Delta tD_0+2\Delta tU(s) D_-+\mathcal{F}_2\left(2\Delta t D_- U(s)\right)\right).
\end{align}
These functions can be bounded if their arguments are $x<1/2$
\begin{align}
        \mathcal{F}_1(x) &\leq 1+x,\\
        \mathcal{F}_2(x) &\leq \frac{x^2}{2}(1+x).
\end{align}

This allows us to bound
\begin{align}
        ||\tH'||&\leq \left[U'(s) D_-+\Delta t D_1 U(s)U'(s) 
                                \left(1+2\Delta t D_- U(s)\right)
                        \right]\nonumber\\\nonumber
                &~~~\times\bigg[1+2\Delta tD_0+2\Delta tU(s) D_-\\
                &~~~~+\frac{\left(2\Delta t D_- U(s)\right)^2}{2}\left(1+2\Delta t D_- U(s)\right)\bigg].
\end{align}
We make the same assumption as Yi and Crosson that \(D_0, \, D_- \in \mathcal O(n)\), \(D_1 \in \mathcal O (n^2)\).  The other essential assumption here is that \(\Delta t \in \mathcal O(n^{-1})\) in order to make the arguments of the $\mathcal{F}$ functions small.  Unfortunately, this assumption is extremely hard to justify in our QAOA setting since we observe that $\Delta t\approx \tau$.  As discussed elsewhere in this paper, $\tau$ is inversely proportional to the spectral gap of the problem, and the spectral gap often scales as an inverse polynomial or exponential in the number of qubits during phase transitions.  At the moment, we will still assume that $\Delta t \in \mathcal O(n^{-1})$, but this is the point in the argument where this method breaks down in our setting.

For now, we continue the argument under the asusmption that $\Delta t \in\mathcal{O}(n^{-1})$ in order to complete the analysis.
With these assumptions, the second term is proportional to \(\mathcal O(1)\) and so we drop it. On the other hand, these assumptions mean that the first term is \(\mathcal O(n)\).

With these assumptions, \(2\Delta t D_- U(s) \le 1/2\) and so
\begin{align}
    ||\tH'||&\leq \left[U'(s) D_-+\frac{3}{2}\Delta t D_1 U(s)U'(s) 
                        \right].
\end{align}

Finally,
\begin{align}
    \epsilon_{\text{tro}} \le& \epsilon'_{\text{adb}} + \epsilon'_{\text{tot}}\\\label{eq:trot_eps}
    \le& \epsilon_{\text{adb}} + \epsilon_{\text{tot}}\\
    =& \mathcal O(\epsilon_{\text{tot}})\\
    =& \mathcal O(\mathcal G(t_f, \tilde H))\\
    \le& \mathcal O \left(\frac{1}{t_f \lambda^3} ||\tilde H'(s)||^2\right)\\
    =& \mathcal O\left(\frac{U'(s) D_-}{t_f \lambda^3}\right) + \mathcal O\left(\frac{(\Delta t D_1 U(s) U'(s))^2}{t_f \lambda^3} \right)\\\label{eq:Yi_final}
    =& \mathcal O\left(\frac{U'(s) D_-}{t_f \lambda^3}\right) + \mathcal O\left(\frac{t_f D_1^2 U(s)^2 U'(s)^2}{p^2 \lambda^3} \right),
\end{align}
where
\begin{align}
    \mathcal G(t_f, \tilde H) =& \frac{1}{t_f}\left( \frac{\|\tilde{H'}(0)\|}{\tilde \lambda(0)^2} + \frac{\|\tilde{H'}(1)\|}{\tilde \lambda(1)^2} \right)\\
    & + \frac{1}{t_f}\int^1_0 \mbox d s \left( \frac{\|\tilde{H''}(s)\|}{\tilde \lambda(s)^2} + 7\frac{\|\tilde{H'}(s)\|^2}{\tilde \lambda(s)^3} \right)\nonumber,
\end{align}
which comes from Ref.~\cite{Jansen} and encapsulates the adiabatic condition.  In the above Eq.~(\ref{eq:trot_eps}) we used 
\begin{align}
\epsilon_{\text{disc}} \equiv& |\epsilon'_{\text{adb}} - \epsilon_{\text{adb}}| \ll \epsilon_{\text{adb}},\\
\tilde \epsilon_{\text{disc}} \equiv& |\epsilon'_{\text{tot}} - \epsilon_{\text{tot}}| \ll \epsilon_{\text{tot}}.
\end{align}

\(\epsilon'_{\text{adb}}\) is the error from a finite time implementation of an adiabatic process, \(\epsilon_{\text{adb}}\), plus discretization error, \(\epsilon_{\text{disc}}\). \(\epsilon'_{\text{tot}}\) is the error from a finite time implementation of an adiabatic process plus discretization error plus Trotter error. \(\epsilon_{\text{tot}}\) is the error in doing the discrete and Trotterized evolution adiabatically. \(\tilde \epsilon_{\text{disc}}\) is the error in discretizing and Trotterizing the adiabatic discretized and Trotterized process.

The dominant term in Eq.~(\ref{eq:Yi_final}) is the second term, and for our purposes, we are interested in how the $U(s)^2 U'(s)^2$ portion effects this as opposed to just taking $U(s)\to 1-s$.  We specifically want to look at the maximum value of this expression as a function of $s$.  For the original result where $U(s)= 1-s$, the maximum value of $U(s)^2 U'(s)^2$ is just one at $s=0$.

In this setting, the enhancement from meshing up with the period of oscillations is ironically not an enhancement so much as a lack of detriment from the oscillations.  
It is easy to see that if $\Delta s = \sigma$, then $U(s) = u_0(s)$ and therefore $U(s)^2U'(s)^2 = u_0(s)^2u_0'(s)^2$, so matching up with the oscillations just makes us to recover the error that would have existed without the oscillations.
On the other hand, not matching up with the oscillations can lead to severe detriments to the error term.

If we choose $\phi = -\frac{\pi \Delta s}{\sigma}$ such that the argument of the oscillations is zero at $s=0$ and then choose the linear ramp $u_0(s) = 1-s$, we can look at the value of this function at $s=0$.  For the linear ramp without the oscillation, $s=0$ is the maximum value of this function, so it will be demonstrative:
\begin{equation}
    U(0)^2U'(0)^2 \to \left(1+\frac{2\pi c_0}{\sigma}\sin \frac{\pi\Delta s}{\sigma}\right)^2.
\end{equation}
For any $\Delta s <\sigma$, the result in a quantity $>1$ and lead to a worsening of the bound.  Even for $\Delta s >\sigma$, there are regions of the curve other than $s=0$ that are detrimentally effected by the oscillations.

For a function other than the linear ramp, the maximum of this quantity could occur somewhere else in $s$.  The only way to ensure that the oscillations will not be deleterious to the Trotterization is to have $\Delta s = \sigma$.

Unfortunately, as stated already, having $\Delta s = \sigma$ violates the assumption that $\Delta t \in\mathcal{O}(n^{-1})$.  It is possible that QAOA angles will start scaling with $n$ differently as we scale up these algorithms, but there is currently no numeric or experimental evidence of this type of scaling.

\subsection{Trotterization Error}
\label{app:trottererror}

Let \(F(t_k, t_{k-1}) = U_1^{(1)}(t_k, t_{k-1}) U_+(t_k, t_{k-1})\).

\begin{equation}
    \frac{\partial F}{\partial t_k} = \frac{\partial U_-^{(+)}}{\partial t_k} U_+ + U_-^{(1)} \frac{\partial U_+}{\partial t_k}.
\end{equation}

\begin{align}
    &\frac{\partial U_-^{(+)}}{\partial t_k} \\
    &= i B(t_k) \exp_-\left(i\int^{t_k}_{t_{k-1}} B(s) \mbox d s\right) \exp_-\left(i\int_{t_{k-1}}^{t_k} A(s) \mbox d s\right) \nonumber\\
    &+ i \exp_-\left(i\int^{t_k}_{t_{k-1}} B(s) \mbox d s\right) A(t_k) \exp_-\left(i\int_{t_{k-1}}^{t_k} A(s) \mbox d s\right) \nonumber.
\end{align}

\begin{equation}
    \frac{\partial U_+}{\partial t_k} = -i(A(t_k) + B(t_k)) U_+.
\end{equation}

\begin{widetext}
\begin{align}
    \frac{\partial F}{\partial t_k} =&  \cancel{i B(t_k) U_-^{(1)} U_+} + \cancel{i A(t_k) U_-^{(1)}U_+} + \left[ \exp_-\left( i \int^{t_k}_{t_{k-1}} B(s) \mbox d s \right), i A(t_k)\right] \exp_-\left( i \int^{t_k}_{t_{k-1}} A(s) \mbox d s \right) \\
    \label{eq:fidelityfirstcancellation}
    &- \cancel{i (A(t_k) + B(t_k)) U_-^{(1)} U_+} - i [U_-^{(1)}, A(t_k) + B(t_k) ] U_+ \nonumber\\
    =& \cancel{i \left[ \exp_-\left( i \int^{t_k}_{t_{k-1}} B(s) \mbox d s \right), A(t_k) \right] \exp_-\left( i\int^{t_k}_{t_{k-1}} A(s) \mbox d s \right) U_+} \\
    \label{eq:fidelitysecondcancellation}
    & - i \exp_-\left(i \int^{t_k}_{t_{k-1}} B(s) \mbox d s \right) \left[ \exp_-\left( i \int^{t_k}_{t_{k-1}} A(s) \mbox d s \right), \cancel{A(t_k)} + B(t_k) \right] U_+ \nonumber\\
    & -i \left[ \exp_-\left( i \int^{t_k}_{t_{k-1}} B(s) \mbox d s \right), \cancel{A(t_k)} + \cancel{B(t_k)} \right] \exp_-\left( i \int^{t_k}_{t_{k-1}} A(s) \mbox d s \right) U_+ \nonumber\\
    =& i \exp_-\left( i \int^{t_k}_{t_{k-1}} B(s) \mbox d s \right) \int^{t_k}_{t_{k-1}} \mbox d u \exp_-\left( i \int^u_{t_{k-1}} A(s) \mbox d s \right) \left[ B(t_k), i A(u) \right] \exp_+\left(-i \int^u_{t_k} A(s) \mbox d s \right) U_+^{(1)} U_-^{(1)} U_+\\
    =& - \exp_-\left( i \int^{t_k}_{t_{k-1}} B(s) \mbox d s \right) \int^{t_k}_{t_{k-1}} \mbox d u \exp_-\left( i \int^u_{t_{k-1}} A(s) \mbox d s \right) \left[ - A(u), B(t_k) \right] \exp_+\left(-i \int^u_{t_k} A(s) \mbox d s \right) \\
    & \times \exp_+\left(-i \int^{t_k}_{t_{k-1}} A(s) \mbox d s \right) \exp_+\left(-i \int^{t_k}_{t_{k_1}} B(s) \mbox d s \right) F \nonumber\\
    =& \exp_-\left( i \int^{t_k}_{t_{k-1}} B(s) \mbox d s \right) \int^{t_k}_{t_{k-1}} \mbox d u \exp_-\left( i \int^u_{t_{k-1}} A(s) \mbox d s \right) \left[ A(u), B(t_k) \right] \exp_+\left(-i \int^u_{t_{k-1}} A(s) \mbox d s \right) \\
    & \times \exp_+\left(-i \int^{t_k}_{t_{k-1}} B(s) \mbox d s \right) F \nonumber
\end{align}
\end{widetext}
where we used the identity
\begin{equation}
    [e^A, B] = -\int^1_0 \mbox d s e^{(1-s)A} [B,A] e^{s A}.
    \label{eq:commutatorofexp}
\end{equation}

Integrating the last equation and using \(F(t_{k-1}, t_{k-1}) = 1\) yields \(F(t_k, t_{k-1}) = \int^{t_k}_{t_{k-1}} C(v, t_{k-1}) F(v, t_{k-1}) \mbox d v\) which produces
\begin{align}
    &\|U_+(t_k, t_{k-1}) - U_+^{(1)}(t_k, t_{k-1}) \|\\
    \le& \int^{t_k}_{t_{k-1}} \mbox d v \| C(v, t_{k-1}) \|\\
    \le& \int^{t_k}_{t_{k-1}} \mbox d v \int^v_{t_{k-1}} \mbox d u \| [A(u), B(v)] \|.
\end{align}

\subsection{Product Formula Perturbations}
\label{app:PF_perturbations}

One important question is whether $\Delta t_k = \tau_k$ is a true minimum or just an enhancement.  In other words, is it beneficial to wiggle slightly away from this.  Furthermore, if it is a true minimum, how much can we wiggle away before messing up the fact that we are in a minimum well.

The results in Section \ref{sec:product} look only at the case where the product formula step size $\Delta t_k$ matches up with the periods of the oscillations $\tau_k$.  What happens if we consider small perturbations such that
\begin{equation}
    \Delta t_k = \tau_k+\epsilon,
\end{equation}
for a small $\epsilon$.

For the purpose of this section, we break the error, $||\hat{U}(0,\Delta t_k)-\hat{U}_{PF}(0,\Delta t_k)||$, up into three portions.  The first is just the base error due to $u_0$ alone without the superposed oscillations.  This section will not consider this portion of the error because we are only concerned with the enhancement due to the oscillations.  The second portion is the cross terms between $u_0(t)$ and the oscillations, which we write as $\mathbb{E}_{CT}$.  In the main portion of the paper, we found that $\mathbb{E}_{CT}=0$.  The last portion, due to just the behavior of the oscillatory portions, is $\mathbb{E}_{Os}$ which is what is responsible for the enhancement we see in the actual results.

Luckily, for both the single oscillation and the aggregate case, simple algebra shows that the cross term error, $\mathbb{E}_{CT}$, cancel out at the first order in $\epsilon$ as well.  Therefore, those terms are consistent with being in an extremum.  For the single oscillation case, the error due to the $\epsilon$ shift is (setting $\phi=0$)
\begin{equation}
 \mathbb{E}_{CT}^{(1)} =  \left|\left|\comm{\hat{B}}{\hat{C}}\right|\right|\int_{-\infty}^{\infty}\!\!\!\!\!d\lambda\frac{i c_k \epsilon ^2 \left(1-e^{2 i \pi  \lambda  \tau _k}\right) \tilde{u}_0(\lambda )}{2 \lambda  \tau_k}+\mathcal{O}(\epsilon^3).
\end{equation}
Here $c_k$ refers to the amplitude of the oscillations during the $k$th time step, and $\tau_k$ is the period of the oscillations during the $k$th time step.

When we look at the error in the cross terms for the entire procedure over multiple oscillations, each with amplitude $c_0$ and period $\tau$, it becomes
\begin{align}
 &\mathbb{E}_{CT} =  \left|\left|\comm{\hat{B}}{\hat{C}}\right|\right|\int_{-\infty}^{\infty}d\lambda
 \bigg(
 \frac{ic_0\epsilon^2\tilde{u}_0(t)}{2\lambda\tau(1-e^{2\pi i \lambda \tau})}\\\nonumber
 &
 \times(e^{2 i \pi  \lambda  \tau }+(2 p-1) e^{2 i \pi  \lambda  (p+1) \tau }-(2 p+1) e^{2 i \pi  \lambda  p \tau }+1)
 \bigg)\\\nonumber
 &+\mathcal{O}(\epsilon^3).
\end{align}

When we consider the case of a single oscillation, the correction from the oscillation, $\mathbb{E}_{Os}$, has no term that is linear in $\epsilon$, meaning that $\Delta t_k = \tau_k$ is an exact minimum. To be exact this term comes out to be (again setting $\phi=0$)
\begin{align}
    \mathbb{E}_{Os}^{(1)}\leq
    \left|\left|\comm{\hat{B}}{\hat{C}}\right|\right|
    \bigg(-\frac{c_k \tau_k^2}{2 \pi }-\frac{\pi  c_k \epsilon ^3}{3 \tau_k}+\mathcal{O}\left(\epsilon ^4\right)\bigg).
\end{align}
This result is mildly problematic because it means that this enhancement is not a minimum here but a higher order critical point.  This problem gets fixed when we consider more oscillations and is a result just of the fact that we are considering a single oscillation here.

To understand how to fix this, we go instead to the aggregate case in the main text where all the oscillations are considered together (under the approximation that the frequency of the oscillations is constant).  There, the expansion results in
\begin{align}
    \mathbb{E}_{Os} = \left|\left|\comm{\hat{B}}{\hat{C}}\right|\right|&\bigg(
    -\frac{c_0 p \tau ^2}{2 \pi }
    \\\nonumber
    &+\frac{1}{6} \pi  c_0 p \left(2 p^2-3 p+1\right) \epsilon ^2
    \bigg)+\mathcal{O}(\epsilon^3).
\end{align}

Note that this is all before coherence or any additional optimization arguments have been made.

\section{Details of Numerics}
\label{app:Jij}

In many of the figures in the main text, we use a specific Ising model.  This Ising model was chosen to have all-to-all couplings with each coupling strength chosen uniformly at random from the range $[-1,1]$.  Other simulations were run in the creation of this paper, and the results were always qualitatively the same.  For completeness here we provide the $J_{ij}$ matrix elements used to generate Figures 
\ref{fig:qaoa_curves}-\ref{fig:eigdist_qaoa} \& \ref{fig:bab_ansatz_example_1}, rounded to three decimal points:
\begin{widetext}
\begin{equation}
    J_{ij}\to\left[
    \begin{array}{cccccccc}
        0 & 0.526 & 0.852 & 0.832 & 0.718 & -0.084 & 0.702 & 0.609\\
        0.526 & 0 & 0.129 & -0.951 & 0.432 & 0.250 & 0.490 & 0.402\\
        0.852 & 0.129 & 0 & 0.243 & 0.708 & -0.648 & 0.753 & -0.743\\
        0.832 & -0.951 & 0.243 & 0 & -0.320 & 0.000 & 0.910 & -0.002\\
        0.718 & 0.432 & 0.708 & -0.320 & 0 & 0.346 & -0.801 & -0.476\\
        -0.084 & 0.250 & -0.648 & 0.000 & 0.346 & 0 & 0.149 & -0.278\\
        0.702 & 0.490 & 0.753 & 0.910 & -0.801 & 0.149 & 0 & 0.509\\
        0.609 & 0.402 & -0.743 & -0.002 & -0.476 & -0.278 & 0.509 & 0\\
    \end{array}
    \right]
\end{equation}
\end{widetext}


\end{document}